\documentclass[amsmath,amssymb]{revtex4}

\usepackage{graphicx}
\usepackage{dcolumn}
\usepackage{bm}
\usepackage{epsfig}
\usepackage{slashed}
\usepackage{subfigure}
\usepackage{titletoc}
\usepackage{hyperref}

\begin{document}
	
	\def\0#1#2{\frac{#1}{#2}}
	\def\bct{\begin{center}} \def\ect{\end{center}}
	\def\beq{\begin{equation}} \def\eeq{\end{equation}}
	\def\bea{\begin{eqnarray}} \def\eea{\end{eqnarray}}
	\def\nnu{\nonumber}
	\def\n{\noindent} \def\pl{\partial}
	\def\g{\gamma}  \def\O{\Omega} \def\e{\varepsilon} \def\o{\omega}
	\def\s{\sigma}  \def\b{c_{B}}  \def\p{\psi} \def\r{\rho}
	\def\G{\Gamma} \def\k{\kappa} \def\l{\lambda} \def\d{\delta} \def\L{\Lambda}
	\def\mrm{\mathrm} \def\pr{\prime} \def\a{\alpha}
	
	\title{Quantum fluctuation energies over a spatially inhomogeneous field background in a chiral soliton model}
	\author{Jiarui Xia$^1$}
    \author{Song Shu$^1$}
	\email {shus@hubu.edu.cn}	
	\author{Xiaogang Li$^{2}$}
	
	\affiliation{1. School of Physics, Hubei
		University, Wuhan, Hubei 430062, China}
	\affiliation{2. Key Laboratory of Quark and Lepton Physics (MOE) and Institute of Particle Physics, Central China Normal University, Wuhan, Hubei 430079, China}
	
	\begin{abstract}
Based on chiral soliton models, the quantum fluctuation energies of quarks over a spatially inhomogeneous meson field background have been thoroughly studied. We have used a systematic calculation scheme initiated by Schwinger, in which the loop quantum fluctuation energies are evaluated by a nontrivial level summation over the eigenvalue spectrum of the effective Hamiltonian of the system. The effective Hamiltonian can be constructed by one loop effective action of fluctuations of quarks over a static chiral soliton field background. The corresponding Dirac equation is obtained. In a static and spatially spherical case and by the hedgehog ansatz the radial part and the angular part of the grand spin of the wave function for the Dirac equation can be separated. Due to the soliton background the eigenvalue spectrum is distorted. The scattering phase shift can be determined by solve the radial equations at different momentum. The density of states in momentum space can be derived. The effective Hamiltonian has been diagonalized in a Hilbert space where the eigenfunctions are labeled by the parity, grand spin and energy. The renormalization scheme can be carried out by a Born subtraction of the phase shift and the compensating Feynman diagram renormalization. Finally the finite quantum fluctuation energies over chiral soliton background at different parities and grand spins have been numerically evaluated, compared and discussed.
	\end{abstract}
	\maketitle
	
	
\section{Introduction}
In recent years there are lots of investigations on the possible inhomogeneous phases in quantum chromodynamics (QCD) systems like the chiral density wave(CDW) phase, the chiral spiral phase, the chiral soliton lattice(CSL) phase and etc.\cite{chen2021,chen2022,FUKUSHIMA201399,BUBALLA201539,Brauner2017csl,PhysRevD.80.074025}. These studies have shown extensive interests in the theoretical exploration about the novel ground states or new phase boundaries in quark/nuclear matter\cite{Ishioka2010cdw,Evans2012baryonic,VOSKRESENSKY2023104030,PhysRevD.103.034030, PhysRevLett.127.232002}. Due to recent advancements in studies of the field of rotating neutron stars and in-depth experimental and theoretical exploration of quark matter at finite densities, the possibility of inhomogeneous phases in quark matter has been considered more seriously\cite{Cardall_2001, PhysRevLett.128.061101,BZDAK20201,PhysRevD.109.036023}. The existence of such inhomogeneous phases is theoretically a semi-classical field configuration formed due to strong coupling effects, resulting in a bound state that creates a spatially inhomogeneous field background, thereby leading to new phases of quark matter on this background. Typically, these semi-classical backgrounds are calculated based on the mean field approximation. However, in the usual discussions of inhomogeneous phases the quantum fluctuation effects have not been systematically considered or rigorously calculated. The calculation of quantum fluctuations on inhomogeneous field backgrounds is very complicated, and even defining a consistent calculation framework within field theory is challenging. Previous studies have used derivative expansion methods, but these studies are given under certain approximations, namely when the spatial variation of the field is relatively slow\cite{PhysRevD.37.1670,PhysRevD.55.3742}. The use of mode truncation methods in finite volumes and ultraviolet truncation methods brings corresponding uncertainties\cite{KAHANA1984462}. In supersymmetric models, the study of quantum corrections in topological soliton background is of particular interest. Due to the supersymmetry, the energy of the soliton will satisfy the Bogomol'nyi-Prasad-Sommerfield (BPS) relation which indicates there is a lower bound of the energy. At the classical level this bound is saturated. At quantum level, since the quantum corrections to the energy and the central charge precisely counterbalance each other, the BPS relations are still saturated to equalities. However, the special results rely on topological soliton background and supersymmetry\cite{PhysRevD.59.045016,DUNNE1999238,GRAHAM1999432,REBHAN2004234,Rebhan2005BPSSO}. Farhi, Graham and Jaffe, among others, have developed a spectral method that can consistently solve for quantum corrections on general soliton backgrounds \cite{farhi1998finite,farhi2002searching,cosmicstring,Graham:2009zz}
. This method is self-consistent and does not introduce additional uncertainties, allowing for the study of quantum fluctuations of topological and non-topological solitons. In this method due to the inhomogeneous background the usual density of states of the momentum in momentum integration of the standard one loop quantum fluctuation energy will receive a correction which can be derived by solving the field equations and evaluating the scattering phase shift in the inhomogeneous background. This method was first proposed by Schwinger in calculations of the electron's quantum fluctuations in certain nontrivial configurations of electromagnetic fields as the background\cite{schwinger1954theory}, and later used by `t Hooft to calculate the quantum corrections of instantons\cite{tHooft1976instanton}, and later developed by Farhi, Graham and Jaffe into an effective method for calculating soliton quantum fluctuations.

This paper focuses on the calculation of quantum fluctuations of chiral solitons. In the past, we have also used the spectral method to calculate the quantum corrections of the Friedberg-Lee (FL) model and extended it to finite temperatures\cite{SHU2017soliton,SHU2021122256,li2022fl}. Now we wish to rigorously calculate the quantum fluctuations of chiral solitons. In the paper by Farhi\cite{farhi2002searching}, the quantum corrections of the 3+1 dimensional chiral Yukawa coupling model have been discussed. In the  paper they have generalized the previously introduced method to the calculation of quantum fluctuations in soliton systems containing chiral fermions, for discussing the stability of fermionic bound states in the electroweak system. We note that in the actual calculations of this literature, when calculating high-order scattering phase shifts, due to the complexity of the calculations, the fake boson substitution method was used, which means that a unified and consistent approach using the Born expansion to calculate phase shifts from the equations at each level was not fulfilled. We hope to use a complete Born expansion and phase shift subtraction scheme to calculate this issue. Moreover, we are using the linear sigma model here to discuss the quantum corrections of the chiral soliton and hope to extend the discussion to the issue of inhomogeneous phases in quark matter in the future.

The organization of this paper is as follows: in section II the chiral soliton model is introduced and the field equations are derived. The semi-classical chiral soliton solution has been calculated. In section III based on a chiral soliton meson field background the one loop energy quantum fluctuations of quark fields have been calculated. The scattering phase shift has been introduced to determine the distorted energy spectrum, which is technically evaluated to the fourth order with a subtle Born subtraction. And the renormalized quantum fluctuation energy has been derived. In section IV, we present the numerical results of momentum dependences of scattering phase shift at different parities and grand spins. The related quantum fluctuation energies have been compared and discussed. The last section is the summary and outlook.

\section{Basic models, equations and ground states}

As we know the QCD Lagrangian density has the chiral symmetry, while at low energies the chiral symmetry of the system is spontaneously or explicitly broken. Considering the difficulties brought by the non-perturbative nature of the QCD itself at low energies, we need extend to the QCD effective models. In our study the linear sigma model will be adopted. It is a chiral model that describes the effective interactions between quarks using scalar fields $\s$ and isospin vector fields $\vec{\pi}$, which can describe the static properties of hadrons through chiral solitons under a semi-classical approximation. In this regard it is also called the chiral soliton model~\cite{BIRSE1984284,diakonov1988chiral,christov1996baryons}.

The Lagrangian density of the chiral soliton model has explicit \(SU(2)_A\) symmetry breaking which form is
\beq
{\cal L}=\bar{\p} \left[i\g^\mu\pl_\mu-g(\s+i\vec{\tau}\cdot\vec{\pi}\g_5)\right] \p + \0{1}{2}\pl_\mu\s\pl^\mu\s + \0{1}{2}\pl_\mu\vec\pi\cdot\pl^\mu\vec\pi - U(\s,\vec{\pi}). \label{Lchiral}
\eeq
The potential of the $\s$ field and the $\vec\pi$ field is:
\beq
U(\s,\vec{\pi}) = \0{\l}{4}(\s^2+\vec{\pi}^2-\nu^2)^2-H\s-\0{m_{\pi}^{4}}{4 \l}+f_{\pi}^{2} m_{\pi}^{2},
\eeq
where $\nu =\sqrt{f_ {\pi}^{2} - \l m_{\pi}^{2}}$. In the Lagrangian density $\p$ represents the quark field which is the two-flavor isospinor of $u, d$ quarks. $\s$ represents the scalar field, and $\vec{\pi}$ represents the isospin vector field $\vec\pi = (\pi_1,\,\pi_2,\,\pi_3) $, which are meson fields. $m_{\pi}$ is $\pi$ meson mass, and $f_ {\pi}$ is $\pi$ meson decay constant.
The parameter $\l$ satisfies the relation $m_\s^2 = m_\pi^2 + 2\l f_\pi^2 $, where $m_\s$ is $\s$ meson mass. The coupling constant $g$ satisfies $m_q=gf_\pi $, where $m_q$ is the mass of the constituent quark in vacuum. $H\s$ is an explicit breaking term. The last two terms in the potential exist to ensure that the minimum potential energy is zero in the absence of quarks in vacuum.

From the least action principle the Euler-Lagrange equations can be derived as
\beq
\pl_\mu \0{\pl {\cal L}}{\pl {\pl_\mu \phi} }-\frac{\pl {\cal L}}{\pl \phi }=0,
\eeq
where the field variables are $\psi$ (or $\bar\psi$), $\s$ and $\vec{\pi}$ components.
In a semi-classical level in order to obtain the ground state of the system, one could take the stationary approximation which means the meson fields $\s(\vec r, t)$ and $\vec{\pi}(\vec r, t)$ are treated as background fields that do not change with time and the temporal part of fermion field can be separated out, which are
\beq
\s(\vec{r},t)=\s(\vec{r}), \ \ \ \vec\pi(\vec{r},t)=\vec{\pi}(\vec{r}),\ \ \ \p(\vec{r},t)=e^{-i \e t} \sum_{i=1}^{N}\Psi_{i}(\vec{r}),
\eeq
where $\e$ is the eigenvalue of energy of the quark. Here the numbers of flavors and colors of quarks are $N_f=2$ and $N_c=3$. One should notice that the quark field has a complicated internal space which will be discussed later. As in our case we have put $N$ valence quarks at the same energy level. In the following the index $i$ of quark field is omitted. In stationary approximation the field equations could be derived as
\begin{gather}\label{fieldeq_chiral}
	\begin{pmatrix}
		g\s & -i\vec\s\cdot\vec\nabla+ig\vec\tau\cdot\vec\pi \\
		-i\vec\s\cdot\vec\nabla-ig\vec\tau\cdot\vec\pi & -g\s
	\end{pmatrix}\Psi = \e\Psi, \\
	-\vec\nabla^2\s+\0{\pl U(\s,\vec{\pi})}{\pl\s}+gN\bar\Psi\Psi=0, \\
	-\vec\nabla^2 \vec{\pi}+\0{\pl U(\s,\vec{\pi})}{\pl\vec{\pi}}+gN\bar\Psi i\g_5\vec{\tau}\Psi=0.
\end{gather}

\subsection{Quark field equations in the hedgehog approximation}

We take the spatial spherical symmetrical case and  introduce the hedgehog approximation
\begin{align}
	&\s(\vec{r})=\s(r), \quad\vec{\pi}(\vec{r})=\hat{\vec{r}} \pi(r),
\end{align}
where $\hat{\vec{r}}$ is the radial unit vector in the real space. After the approximation, the $\s(r),\pi (r)$ fields are only functions of the radial distance $r$, and the orientation of the $\vec\pi(r)$ field in the isospin space is aligned with the direction of the radial unit vector $\hat{\vec{r}}$ in the real space. The quark field $\Psi(\vec r)$ has a multiple internal space, which includes flavor, color and spin spaces. In our case the quark wave functions are degenerate in the color space. Thus the internal space of quark field is a direct product isospin space of flavor and spin. In order to simplify the quark field equation \eqref{fieldeq_chiral} and separate the spatial radial part of the quark wave function, one should project the wave function onto the right bases of eigen spinors. Following the method in references \cite{KAHANA1984462,farhi2002searching,diakonov1988chiral} we construct spinors that are eigenstates of parity and total grand spin, where the grand spin $\vec G$ is defined as the sum of orbital angular momentum $\vec {L}$, spin $\vec {S}$ and isospin $\0{1}{2}\vec {\tau}$. According to the angular momentum coupling theory there are:
\begin{eqnarray}
	\vec{G} & = & \vec{L}+\vec{S}+\frac{1}{2}\vec{\tau}, \\
	\vec{J} & = & \vec{L}+\vec{S},
\end{eqnarray}
where $\vec{J}$ is the usual total angular momentum.
We assume $G,j,l,\Pi$ as the quantum numbers of the grand spin, total angular momentum, orbital angular momentum and parity. For the quark eigenstate with given $G$ and $\Pi$ the corresponding wave function can be labeled as $\Psi_{G,\Pi}$. For the channels with parity $\Pi=+(-1)^G$ the wave function can be constructed as:
\beq
\Psi_{G,\Pi=+}(\vec{r})=
\begin{pmatrix}
	ig_1(r)y_{G+\tfrac{1}{2},G}(\hat{\vec{r}}) \\
	f_1(r)y_{G+\tfrac{1}{2},G+1}(\hat{\vec{r}})
\end{pmatrix}+
\begin{pmatrix}
	ig_2(r)y_{G-\tfrac{1}{2},G}(\hat{\vec{r}}) \\
	-f_2(r)y_{G-\tfrac{1}{2},G-1}(\hat{\vec{r}})
\end{pmatrix},
\eeq
where the spherical harmonic functions $y_{j, l}(\hat{\vec{r}})$ with $j = G\pm\0 {1} {2}$ and $l=j\pm\0{1}{2}$ are the two-component spinors in both spin and isospin space. $f_i(r)$ and $g_i(r)$ are the radial functions. For the opposite parity $\Pi=-(-1)^G$ the wave function is the following form
\beq
\Psi_{G,\Pi=-}(\vec{r})=
\begin{pmatrix}
	ig_1(r)y_{G+\tfrac{1}{2},G+1}(\hat{\vec{r}}) \\
	-f_1(r)y_{G+\tfrac{1}{2},G}(\hat{\vec{r}})
\end{pmatrix}+
\begin{pmatrix}
	ig_2(r)y_{G-\tfrac{1}{2},G-1}(\hat{\vec{r}}) \\
	f_2(r)y_{G-\tfrac{1}{2},G}(\hat{\vec{r}})
\end{pmatrix}.
\eeq
Substituting the above wave functions into the quark field equation \eqref{fieldeq_chiral} and considering the hedgehog approximation, after a tedious calculation one can derive the coupled first-order differential equations for the radial functions $f_i(r)$ and $g_i(r)$. For the parity $\Pi=+(-1)^G$ the radial equations are:
\beq
\left\{\begin{array}{l}
	f_1'+\dfrac{G+2}{r}f_1 - (\e-g\s)g_1 + \dfrac{g\pi}{2G+1} \left[f_1+2\sqrt{G(G+1)}f_2\right]=0 \\
	f_2'-\dfrac{G-1}{r}f_2 + (\e-g\s)g_2 - \dfrac{g\pi}{2G+1} \left[f_2-2\sqrt{G(G+1)}f_1\right]=0 \\
	g_1'-\dfrac{G}{r}g_1 + (\e+g\s)f_1 - \dfrac{g\pi}{2G+1} \left[g_1-2\sqrt{G(G+1)}g_2\right]=0 \\
	g_2'+\dfrac{G+1}{r}g_2 - (\e+g\s)f_2 + \dfrac{g\pi}{2G+1} \left[g_2+2\sqrt{G(G+1)}g_1\right]=0
\end{array}\right.,\label{p-G}
\eeq
where the prime indicates the first derivative with respect to $r$. For the parity $\Pi=-(-1)^G$ the radial equations are:
\beq
\left\{\begin{array}{l}
	g_1'+\dfrac{G+2}{r}g_1 - (\e+g\s)f_1 - \dfrac{g\pi}{2G+1} \left[g_1-2\sqrt{G(G+1)}g_2\right]=0 \\
	g_2'-\dfrac{G-1}{r}g_2 + (\e+g\s)f_2 + \dfrac{g\pi}{2G+1} \left[g_2+2\sqrt{G(G+1)}g_1\right]=0 \\
	f_1'-\dfrac{G}{r}f_1 + (\e-g\s)g_1 + \dfrac{g\pi}{2G+1} \left[f_1+2\sqrt{G(G+1)}f_2\right]=0 \\
	f_2'+\dfrac{G+1}{r}f_2 - (\e-g\s)g_2 - \dfrac{g\pi}{2G+1} \left[f_2-2\sqrt{G(G+1)}f_1\right]=0
\end{array}\right..\label{n-G}
\eeq

Now let us consider the case of $G=0$ alone. In this case, $f_2(r)=g_2(r)= 0$. The wave functions are in the following forms for positive and negative parities respectively,
\beq
\Psi_{G=0,\Pi=+}(\vec{r})=
\begin{pmatrix}
	ig_1(r)y_{\tfrac{1}{2},0}(\hat{\vec{r}}) \\
	f_1(r)y_{\tfrac{1}{2},1}(\hat{\vec{r}})
\end{pmatrix}, \ \ \ \Psi_{G=0,\Pi=-}(\vec{r})=
\begin{pmatrix}
	ig_1(r)y_{\tfrac{1}{2},1}(\hat{\vec{r}}) \\
	-f_1(r)y_{\tfrac{1}{2},0}(\hat{\vec{r}})
\end{pmatrix}.
\eeq
The above first-order radial equations will be simplified accordingly. For the positive parity the radial equations are
\beq
\left\{\begin{array}{l}
	f_1'+\dfrac{2}{r}f_1 - (\e-g\s)g_1 + g\pi f_1=0, \\
	
	g_1' + (\e+g\s)f_1 - g\pi g_1=0.
\end{array}\right.
\eeq
For the negative parity the radial equations are
\beq
\left\{\begin{array}{l}
	g_1'+\dfrac{2}{r}g_1 - (\e+g\s)f_1 - g\pi g_1=0, \\
	f_1'+ (\e-g\s)g_1 + g\pi f_1=0.
\end{array}\right. \label{eq_fg0}
\eeq

\subsection{Ground states and chiral solitons}
In order to discuss the ground state solution, we combine the quark field equations for the case of $G=0$ with the radial equations of $\s$ and $\pi$ meson fields, and one can obtain the ground state equations for both positive and negative parities. However, by numerical evaluation and analysis, we find that the bound state solution can only be obtained with positive parity for the $G=0$ case. In the channel of positive parity the coupled equations for the ground state are:
\begin{align}
	& \0{{\rm d} f_1(r)}{{\rm d} r}=-\0{2}{r}f_1(r)+(\e-g \s(r))g_1(r)-g \pi(r)f_1(r), \\
	& \0{{\rm d} g_1(r)}{{\rm d} r}=-(\e+g \s(r)) f_1(r)+g \pi(r) g_1(r), \\
	& \0{\pl U}{\pl \s}=\0{{\rm d}^{2} \s(r)}{{\rm d} r^{2}}+\0{2}{r} \0{{\rm d} \s(r)}{{\rm d} r}-N g\left(g_1^{2}(r)-f_1^{2}(r)\right), \\
	& \0{\pl U}{\pl \pi}=\0{{\rm d}^{2} \pi(r)}{{\rm d} r^{2}}+\0{2}{r} \0{{\rm d} \pi(r)}{{\rm d} r}-\0{2 \pi(r)}{r^{2}}-2 N g f_1(r) g_1(r).
\end{align}
These coupled equations could be numerically solved under the boundary conditions
\begin{align}
	& f_1(0)=0, \frac{\mrm{d} \s(0)}{\mrm{d}r}=0,\,\pi(0)=0 \\
	& g_1(\infty)=0,\, \s(\infty)=f_{\pi},\,\pi(\infty)=0
\end{align}
together with quark field normalization condition
\begin{align}
& 4 \pi \int r^{2}\left(f_1^{2}(r)+g_1^{2}(r)\right) \mrm{d} r=1.
\end{align}

\begin{figure}[htbp]
	\centering
	\begin{minipage}{0.8\textwidth}
		\centering
		\includegraphics[width=\textwidth]{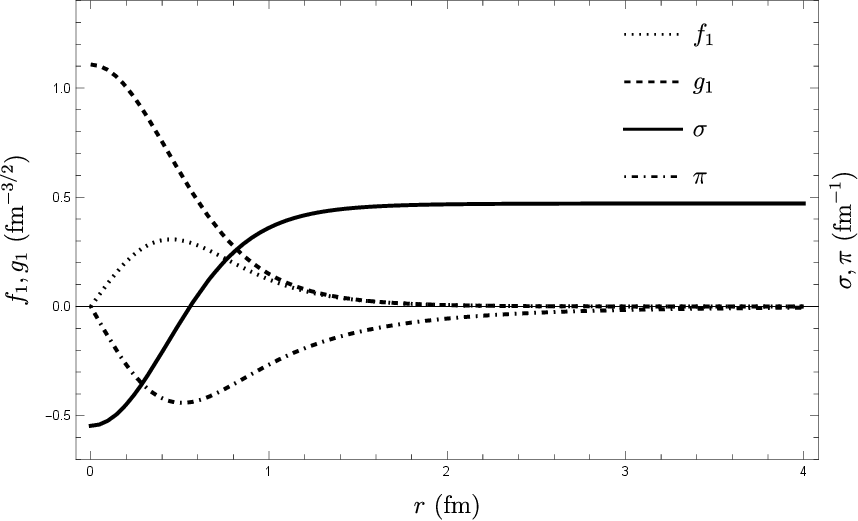}
	\end{minipage}\\
	\vspace{0.3cm}
	\caption{The solutions $f_{1}(r),\,g_{1}(r)$, $\s (r)$ and $\pi (r)$ of the chiral soliton equations.}\label{solu_chiral}
\end{figure}

For a chiral baryon there are $N=3$ quarks at energy level $\e$.
The model parameters used in this study are: $\pi$ meson decay constant $f _\pi = 93 {\rm MeV}$, $\pi$ meson mass $m _\pi = 138{\rm MeV}$, the coupling constant $g\approx 5.28 $ and $\l\approx 82.1$, $\s$ meson mass $m _\sigma = 1200{\rm MeV}$ and quark mass $m_q = 491{\rm MeV}$~\cite{christov1996baryons,BIRSE1984284}. After the numerical calculation, the $f_1 (r)$, $g_1 (r)$, $\s(r)$ and $\pi(r)$ can be numerically evaluated, and the results are shown in the Fig.\ref{solu_chiral}. The energy eigenvalue of the valence quark could be also determined as $\e\approx 0.216\, {\rm fm}^{-1}\approx 42.61{\rm MeV}$. \\

The classical energy of the chiral soliton can be expressed as
\beq
E_{\rm cl}=3\e+4\pi\int_0^\infty {\rm d}r\,r^2\left[\0{1}{2}\left(\0{{\rm d} \s}{{\rm d} r}\right)^2+\0{1}{2}\left(\0{{\rm d} \pi}{{\rm d} r}\right)^2+\0{\pi^2}{r^2}+U(\s,\pi)\right],
\eeq
which is numerically evaluated as $E_ {\rm cl}\approx 5.83\, {\rm fm}^{-1}\approx 1149.8\, {\rm MeV} $. In previous studies the chiral soliton in this model has been interpreted as a baryon. And here our classical chiral soliton results are consistent with those in previous literatures.

\section{Quantum fluctuation energies over inhomogeneous background}
\subsection{One loop quantum fluctuation energy and renormalization}
Now let us discuss quantum fluctuations over the classical soliton background. In our case the chiral symmetry is explicitly breaking. The system has a a unique vacuum state at $\pi_v = 0$, $\s_v = f_\pi$. The meson fields can be decomposed into two parts: the homogeneous vacuum part and the inhomogeneous ground state background part as
\beq
\s(r)=\s_v+\tilde{\s}(r),\pi(r)=\pi_v+\tilde{\pi}(r) \label{bag}
\eeq

In principle, quantum fluctuations should include loop corrections from both the quark field and the meson fields ($\s$, $\pi$). However, if one consider them together, it will be very difficult. The meson fluctuations in this model have been studied separately in the previous literature\cite{PhysRevC.70.015204,WENG2025139587}. In this work we mainly focus on studying the quantum fluctuations from the quark field. As in the inhomogeneous background the Dirac sea of quarks has been polarized. One loop quantum fluctuation from the sea quarks will be calculated in the inhomogeneous background.
The corresponding action of sea quarks in the background field $\tilde{\s}(r)$, $\tilde{\pi}(r)$ is
\beq
S_{\p}[\tilde{\s}(r),\tilde{\pi}(r)]=\int {\rm d}^{4} x\left[\bar{\p}\left(i \g_{\mu} \pl^{\mu}-m_{q}-g(\tilde{\s}+i\g_5\vec{\tau}\cdot \hat{\vec{r}} \tilde{\pi})\right) \p+{\cal L}_{\rm ct}\right],
\eeq
where
\beq
{\cal L}_{\rm ct}=2a ((\pl_{\mu} \s)^2+(\pl_{\mu} \pi)^2)-2b(\s^2+\vec{\pi}^2-\s_v^2)-4c(\s^2+\vec{\pi}^2-\s_v^2)^2.
\eeq
${\cal L}_{\rm ct}$ represents the renormalization counter term, where $a$, $b$, and $c$ are the cut-off dependent constants. By using the dimensional regularization one can obtain:
\beq
\begin{aligned}
	a & =-\frac{g^2}{(4 \pi)^2}\left\{\frac{1}{\epsilon}-\gamma-\frac{2}{3}+\ln \left(\frac{4 \pi \mu^2}{m_{q}^2}\right)-6 \int_0^1 d x x(1-x) \ln \left[1-x(1-x) \frac{m_s^2}{m_{q}^2}\right]\right\}, \\
	b & =-\frac{g^2 m_{q}^2}{(4 \pi)^2}\left\{\frac{1}{\epsilon}-\gamma+1+\ln \left(\frac{4 \pi \mu^2}{m_{q}^2}\right)\right\},\\
	c &=-\frac{g^4}{(4\pi)^2} \left\{\frac{1}{\epsilon}-\gamma+\ln \left(\frac{4 \pi \mu^2}{m_{q}^2}\right)-\frac{m_{s}^2}{4 m_{q}^2}-\frac{3}{2} \int_0^1 d x \ln \left[1-x(1-x) \frac{m_s^2}{m_{q}^2}\right]\right\},
\end{aligned}
\eeq
where $d=4-2\epsilon$, $m_s =\sqrt{\l}f _\pi$, and $\mu$ is the scale parameter introduced to make the coupling constant $g$ dimensionless~\cite{farhi2002searching}. The effective action can be defined as
\beq
e^{i S_{\rm eff}[\tilde{\s},\tilde{\pi}]}=\0{\int[D \p]\left[i D \p^{\dagger}\right] e^{i S_{\p}[\tilde{\s},\tilde{\pi}]}}{\int[D \p]\left[i D \p^{\dagger}\right] e^{\left.i S_{\p}[\tilde{\s}]\right|_{\tilde{\s}=\tilde{\pi}=0}}}.
\eeq
After integrating the $\p$ field, the one loop effective action can be written as
\beq
S_{\rm eff}[\tilde{\s},\tilde{\pi}]=\operatorname{Tr} \log \0{h_{D}[\tilde{\s},\tilde{\pi}]}{h_{D}[\tilde{\s}] \mid_{\tilde{\s}=\tilde{\pi}=0}}+S_{\rm ct}, \label{Seff}
\eeq
where $h_{D}$ is the Dirac operator in the form of $h_{D} = i\g_{\mu}\pl^{\mu} -m_{q} -g (\tilde {\s} + i\vec {\tau}\cdot\hat{\vec {r}}\tilde {\pi}) $, and $S_{\rm ct}$ is the counter term. The effective one loop vacuum energy can be obtained from $E_{\rm vac}=-S_{\rm eff}\int{\rm d}t$, and the result is
\beq
E_{\rm vac}[\tilde{\s},\tilde{\pi}]=E_{\rm vac}^{\p}[\tilde{\s},\tilde{\pi}]+E_{\rm ct}[\tilde{\s},\tilde{\pi}],
\eeq
where $E_{\rm ct} [\tilde {\s},\tilde {\pi}]$ is the corresponding energy renormalization counter term, and
$E_{\rm vac}^{\p}$ is the difference between the energy of the Dirac sea filled with quarks with a background field $\tilde{\s}(r)$, $\tilde{\pi}(r)$ and the energy of the Dirac sea filled with quarks without the background field, which can be expressed as
\beq
E_{\rm vac}^{\p}[\tilde{\s},\tilde{\pi}]=-\frac{1}{2}\left[\sum_{n}(2G_{n}+1)|E_{n}|+\sum_{\e_{\eta}}|E_{\eta}(k)|-\sum_{\e_{k}}|E_{q}(k)|\right], \label{Evac_dffrce}
\eeq
where $E_q(k)=\omega\sqrt{k^2 + m_q^2}$ is the continuous energy spectrum of the free quark field, and $\omega$ denotes the sign of the energy eigenvalue. $E_n$ and $E_\eta$ are all the discrete and continuous eigenvalues of the following stationary Dirac equations:
\beq
H_{\rm eff}(\tilde{\s},\tilde{\pi})\p_{\a}=\left[-i\g_{0}\vec\g \cdot\nabla+\g_{0}m_{q}+\g_{0}g(\tilde{\s}+i\g_5\vec{\tau}\cdot\hat{\vec{r}}\tilde{\pi})\right]\p_{\a}=E_{\a} \p_{\a}, \label{Dirac_eq}
\eeq
where $H_{\rm eff}$ is an effective hamiltonian of quarks with a static background field of chiral soliton. By the partial wave expansion method, the sum of momentum $k$ can be expressed as the following integral:
\bea
\sum_{\e_{k}}|E_{q}(k)|=\sum_{G,\omega,\Pi}(2 G+1)\int {\rm d}k \rho_{G,\omega,\Pi}^{\rm free}(k)|E_{q}(k)|,  \\
\sum_{\e_{\eta}}|E_{\eta}(k)|=\sum_{G,\omega,\Pi}(2 G+1)\int {\rm d}k \rho_{G,\omega,\Pi}(k)|E_{q}(k)|
\label{sum-k}
\eea
where $\rho_{G,\omega,\Pi}^{\rm free}(k)$ and $\rho_{G,\omega,\Pi}(k)$ are the spectral density of states in momentum space in free case without background and in scattering case with background respectively. Notice that $G$ and $\Pi$ are quantum numbers of the grand spin and the parity. The difference between the two functions of density of states is proportional to the derivative of the scattering phase shift:
\beq
\rho_{G,\omega,\Pi}(k)-\rho_{G,\omega,\Pi}^{\rm free}(k)=\0{1}{\pi} \0{{\rm d}\d_{G,\omega,\Pi}(k)}{{\rm d}k},
\label{sum-k}
\eeq
where $\d_{G,\omega,\Pi}(k)$ is the scattering phase shift in the channel with given $G$, $\omega$ and $\Pi$. So $E_{\rm vac}^{\p}$ can be reexpressed as an integral on the scattering phase shift:
\beq
E_{\rm vac}^{\p}[\tilde{\s},\tilde{\pi}]=-\frac{1}{2}\sum_{n} (2G_{n}+1)|E_{n}|-\frac{1}{2}\sum_{G,\omega,\Pi}(2 G+1) \int {\rm d}k \0{1}{\pi} \0{{\rm d}\d_{G,\omega,\Pi}(k)}{{\rm d}k} |E_{q}(k)|. \label{Evac_sum_int}
\eeq
Since the asymptotic behavior of $\d_ {G,\omega,\Pi}(k)$ as $k\rightarrow\infty$ is $\0{1}{k}$, it is the logarithmic divergence. Additionally there are the sums over $G$, $\omega$ and $\Pi$, thus $E_ {\rm vac}^{\p}$ is highly divergent. In order to make it finite, one should adopt a systematic renormalization scheme. Here we have been following the way used in the literatures \cite{Graham:2009zz, farhi1998finite, cosmicstring, farhi2002searching}. As for high momentum, the Born expansion becomes a good approximation to the phase shift. So by subtracting the successive terms in the Born series to a certain order from the phase shift, the divergences could be removed. The regularized scattering phase shift has the following form
\beq
\bar{\d}_{G,\omega,\Pi}(k)\equiv \d_{G,\omega,\Pi}(k)-\sum_{\ell=1}^{n}\d_{G,\omega,\Pi}^{(\ell)}(k), \label{subtracted_fs}
\eeq
where $\d_{G,\omega,\Pi}^{(\ell)}(k)$ is the $\ell th$ Born approximation to $\d_ {G,\omega,\Pi}(k)$. $\bar{\d}_{G,\omega,\Pi}(k)$ is also called the subtracted phase shift. However if one subtracts a term in the Born expansion, one needs to add back the equivalent Feynman diagram. The corresponding divergent Feynman diagrams should be properly renormalized~\cite{farhi1998finite,farhi2002searching,cosmicstring}. Finally one has the finite one loop result of the vacuum fluctuation energy as
\beq
E_{\rm vac}^{\rm ren}[\tilde{\s},\tilde{\pi}]=-\frac{1}{2}\sum_{n}(2G_{n}+1)|E_{n}|-\frac{1}{2}\sum_{G,\omega,\Pi}(2 G+1) \int {\rm d}k \0{1}{\pi} \0{{\rm d}\bar{\d}_{G,\omega,\Pi}(k)}{{\rm d}k} |E_{q}(k)|+\Gamma_2+\Gamma_4, \label{Evac_sum_int}
\eeq
where for the subtracted phase shift when $G=0$, the Born subtraction needs to be evaluated to the second order; while for $G\neq0$ case the Born subtraction needs to be evaluated to the fourth order. The phase shifts and Born subtraction of the phase shifts can be calculated by the radial equations, which will be discussed in the next section. $\Gamma_2$ and $\Gamma_4$ in (\ref{Evac_sum_int}) are finite terms evaluated from the Feynman diagrams. The quadratic divergent fermion diagrams have been renormalized by the corresponding counter terms in the usual way and the result is

\begin{align}
	\Gamma_2= & \frac{g^2}{\pi^2} \int_0^{\infty} d q q^2\left[s^2(q)+p^2(q)\right]
	\left\{q^2+m_{s}^{2}-6 \int_0^1 d x\left[m_{q}^2+x(1-x) q^2\right] \ln \frac{m_{q}^2+x(1-x) q^2}{m_{q}^2-x(1-x) m_{s}^{2}}\right\} \notag\\
	- & \frac{g^2}{\pi^2} \int_0^{\infty} d q q^2 p^2(q)\left\{m_{s}^{2}+2 m_{q}^2 \int_0^1 d x\left[3 \ln \left(1-x(1-x) \frac{m_{s}^{2}}{m_{q}^2}\right)\right.\right.
	\left.\left.-2 \ln \left(1+x(1-x) \frac{q^2}{m_{q}^2}\right)\right]\right\},
\end{align}
and the Fourier transform is introduced:
\beq
s(q)=\int_{0}^{\infty} d r r^{2} \frac{\sin q r}{q r}\tilde\s(r),  \quad
p(q)=\int_{0}^{\infty} d r r^{2}\left[\frac{\cos q r}{q r}-\frac{\sin q r}{(q r)^{2}}\right] \tilde \pi(r).
\eeq
However, the renormalization of the logarithmic divergence involves a nontrivial calculation. To maintain consistency between the renormalization scheme and the current Born subtraction scheme, we follow the fake boson approach as described in Refs.~\cite{farhi2002searching,cosmicstring}. The key step is to add back the corresponding logarithmically divergent Feynman diagram and renormalize it. Details of the renormalization procedure using the fake boson method are given in Appendix A. After renormalization, the resulting fake boson Feynman diagram contributes the finite term
\begin{gather}
\Gamma _{4} =-c_{B} \int_0^{\infty} \frac{dqq^2}{2\pi ^2}\tilde{V}_{B}(q)\tilde{V}_{B}(-q)
\left[\int_{0}^{1}dx\frac{x(1-x)}{m_B^2-x(1-x)m_f^2}(q^2+m_f^2)+\int_{0}^{1}dx\ln\frac{m_B^2+x(1-x)q^2}{m_B^2-x(1-x)m_f^2} \right],
\end{gather}
where
\begin{gather}
\tilde{V}_{B}(q) =\int_{0}^{\infty} d r r^{2} \frac{\sin q r}{q r}V_{B}(r).
\end{gather}
$m_B$ and $V_{B}(r)$ denote the mass and background potential of the fake boson, respectively. The scale factor $c_B$ is determined by matching the logarithmically divergent energy from the third- and fourth-order fermion phase shifts with the corresponding energy from the second-order fake boson phase shift. Meanwhile, $m_f$ represents the assigned mass of the external line in the boson loop, which is fixed by matching the renormalization energies in the two fake boson schemes. Further discussions and detailed calculations are provided in Appendix A.

\subsection{Evaluations of scattering phase shifts}
Next, based on the background fields $\s(r)$ and $\pi(r)$ we can calculate the scattering phase shift from the quark fluctuation field radial equations under different quantum numbers of the grand spin $G$ and the parity $\Pi$. In the following we take the case of parity $\Pi=-(-1)^G$ as an example to briefly illustrate the calculation process of the scattering phase shift at $G=0$ and $G\neq0$ separately. In the following discussion of this section the labels for $G$, $\omega$ and $\Pi$ will be suppressed.
\subsubsection{$G=0$ case}
When $G = 0 $, the equations of negative parity are
\beq
\left\{\begin{array}{l}
	g_1'+\dfrac{2}{r}g_1 - (\e+g\s)f_1 - g\pi g_1=0, \\
	f_1'+ (\e-g\s)g_1 + g\pi f_1=0,
\end{array}\right.
\eeq
At $r\to\infty $, considering $\s(r)\to\s_v,\,\pi\to 0 $, the asymptotic forms of the equations \eqref{eq_fg0} are
\beq
\left\{\begin{array}{l}
	g_1'+\dfrac{2}{r}g_1 - (\e+g\s_v)f_1=0, \\
	f_1'+ (\e-g\s_v)g_1=0,
\end{array}\right.
\eeq
From the characteristics of the equations, combined with the recursive relationship satisfied by the spherical Hankel function, the form of solutions of scattering outgoing wave can be set and the boundary conditions can be given, which are
\begin{align}
	f_1(r) &= u(r)h_0(kr),\,u(r)\to k/(\e+g \s_v)\,(r\to\infty),\\
	g_1(r) &= v(r)h_1(kr),\,v(r)\to 1\,(r\to\infty),
\end{align}
where $h _\nu (x)$ is the spherical Hankel function of the first kind, and $k =\sqrt {\e^2-g^2\s_v^2} $. Substituting the above undetermined formulas into the equation \eqref {eq_fg0}, one obtains the equations that $u(r)$ and $v(r)$ satisfy
\beq
\left\{\begin{array}{l}
	v' h_1+v\left(h_1'+\dfrac{2}{r}h_1-g\pi h_1\right)-u(g\s+\e)h_0=0, \\
	u' h_0+u(h_0'+g\pi h_0)-v(g\s-\e)h_1=0,
\end{array}\right. \label{chiral_uv0}
\eeq
where the notation prime represents the derivative to $r$. Combining the above equations with the boundary conditions at $r\to\infty$, one can numerically solve the functions $u(r)$ and $v(r)$. Usually the upper part $u(r)$ and lower part $v(r)$ of the Dirac spinor could be decoupled. The scattering phase shift can be obtained from either $u(r)$ or $v(r)$. As the general scattering solution is constructed by a superposition of an incoming wave and an outgoing wave in the spherical potential background, the phase shift is understood as the phase difference between the scattering outgoing wave function with the background and the free outgoing wave function without the background at $r\to\infty$. Considering this and imposing that the scattering solution be regular at $r\to 0$, the phase shift could be obtained at $G=0$ and given $k$ as
\beq
\d_0(k)=\0{1}{2i}\lim_{r\to 0}\ln(u^*(r)u^{-1}(r))=\0{1}{2i}\lim_{r\to 0}\ln(v^*(r)v^{-1}(r))
\eeq
\subsubsection{$G\neq 0$ case}
For the case of $G\neq 0$ and the parity $\Pi=-(-1)^G$, the solution has a matrix form. Usually the first order differential equations should be decoupled to two sets of the second order equations of the upper part and the lower part of the Dirac spinor. Each part is still a two flavor isospinor. For the upper part the radial wave functions are $g_1(r)$ and $g_2(r)$, while for the lower part they are $f_1(r)$ and $f_2(r)$. As it is a two-channel scattering problem in either the upper or lower part, the two linearly independent scattering boundary conditions should be implemented to each part. One can construct the matrix forms of solutions for the upper part and lower part separately, which are
\beq
\mathcal{G}(r)\equiv\begin{pmatrix}
	g^{(1)}_1(r) & g^{(2)}_1(r) \\
	g^{(1)}_2(r) & g^{(2)}_2(r)
\end{pmatrix}=H_v(kr)V(r), \ \ \
\mathcal{F}(r)\equiv\begin{pmatrix}
	f^{(1)}_1(r) & f^{(2)}_1(r) \\
	f^{(1)}_2(r) & f^{(2)}_2(r)
\end{pmatrix}=H_u(kr)U(r),
\eeq
where the two boundary conditions are labeled by the upper indices $j$ in $g^{(j)}_i$ (or $f^{(j)}_i$). In the second equality the matrix form solution has been modified to a multiplicative matrix form relative to the matrix solution to the free differential equations without the background, and
\begin{gather}
	V(r) \equiv\left(\begin{array}{ll}
		v_{1}^{(1)}(r) & v_{1}^{(2)}(r) \\
		v_{2}^{(1)}(r) & v_{2}^{(2)}(r)
	\end{array}\right),\quad
	H_v(k r) \equiv\left(\begin{array}{cc}
		h_{G+1}(k r) & 0 \\
		0 & h_{G-1}(k r)
	\end{array}\right), \label{vh}\\	
U(r) \equiv\left(\begin{array}{ll}
		u_{1}^{(1)}(r) & u_{1}^{(2)}(r) \\
		u_{2}^{(1)}(r) & u_{2}^{(2)}(r)
	\end{array}\right),\quad
	H_u(k r) \equiv\left(\begin{array}{cc}
		h_{G}(k r) & 0 \\
		0 & h_{G}(k r)
	\end{array}\right),  \label{uh}
\end{gather}
where $h_n$ is the spherical Hankel function with $n=G-1,G,G+1$. In the following we have adopted a strategy which is a little different from that in Farhi's work~\cite{farhi2002searching}. We do not decouple the first order equations of the upper part and lower part of the Dirac spinor to the second order differential equations. Instead we solve the first order coupled differential equations as a whole. The specific calculation of the Born subtraction of the phase shift by the first order differential equations is presented in Appendix B. The reason for choosing this strategy is that it is more convenient and efficient in the later calculations of Born subtractions of the phase shifts to higher orders. However, in order to check the results calculated by the first order differential equations, we also carry out the calculations by the second order differential equations which are presented in Appendix C, where the phase shifts and corresponding Born subtraction are also evaluated by the second order differential equations. In Farhi's work~\cite{farhi2002searching}, the Born subtraction of the phase shift has been only calculated to the second order, while the higher order Born subtraction has been substituted by the fake boson subtraction. In Appendix D, we have made a comparison between the results of the total energies evaluated by the Born subtraction and the fake boson subtraction schemes.

Substituting the matrix solutions into the first order equations \eqref{n-G}, one can obtain the following differential equations of matrix elements of $V(r)$ and $U(r)$ for the two boundary conditions separately, which are
\beq
\left\{\begin{aligned}
	& v_1^{(1) \prime}+v_1^{(1)} \frac{h_{G+1}^{\prime}}{h_{G+1}}+\frac{G+2}{r} v_1^{(1)}-(\e+g \sigma) \frac{h_G}{h_{G+1}} u_1^{(1)}-\frac{g \pi}{1+2 G}\left(v_1^{(1)}-2 \sqrt{G(1+G)} \frac{h_{G-1}}{h_{G+1}} v_2^{(1)}\right)=0 ; \\
	& v_2^{(1) \prime}+v_2^{(1)} \frac{h_{G-1}^{\prime}}{h_{G-1}}-\frac{G-1}{r} v_2^{(1)}+(\e+g \sigma) \frac{h_G}{h_{G-1}} u_2^{(1)}+\frac{g \pi}{1+2 G}\left(v_2^{(1)}+2 \sqrt{G(1+G)} \frac{h_{G+1}}{h_{G-1}} v_1^{(1)}\right)=0 ; \\
	& u_1^{(1) \prime}+u_1^{(1)} \frac{h_G^{\prime}}{h_G}-\frac{G}{r} u_1^{(1)}+(\e-g \sigma) \frac{h_{G+1}}{h_G} v_1^{(1)}+\frac{g \pi}{1+2 G}\left(u_1^{(1)}+2 \sqrt{G(1+G)} u_2^{(1)}\right)=0 ; \\
	& u_2^{(1) \prime}+u_2^{(1)} \frac{h_G^{\prime}}{h_G}+\frac{G+1}{r} u_2^{(1)}-(\e-g \sigma) \frac{h_{G-1}}{h_G} v_2^{(1)}-\frac{g \pi}{1+2 G}\left(u_2^{(1)}-2 \sqrt{G(1+G)} u_1^{(1)}\right)=0 ;
\end{aligned}\right.\label{n-G-I}
\eeq
and
\beq
\left\{\begin{aligned}
	& v_1^{(2) \prime}+v_1^{(2)} \frac{h_{G+1}^{\prime}}{h_{G+1}}+\frac{G+2}{r} v_1^{(2)}-(\e+g \sigma) \frac{h_G}{h_{G+1}} u_1^{(2)}-\frac{g \pi}{1+2 G}\left(v_1^{(2)}-2 \sqrt{G(1+G)} \frac{h_{G-1}}{h_{G+1}} v_2^{(2)}\right)=0 ; \\
	& v_2^{(2) \prime}+v_2^{(2)} \frac{h_{G-1}^{\prime}}{h_{G-1}}-\frac{G-1}{r} v_2^{(2)}+(\e+g \sigma) \frac{h_G}{h_{G-1}} u_2^{(2)}+\frac{g \pi}{1+2 G}\left(v_2^{(2)}+2 \sqrt{G(1+G)} \frac{h_{G+1}}{h_{G-1}} v_1^{(2)}\right)=0 ; \\
	& u_1^{(2) \prime}+u_1^{(2)} \frac{h_G^{\prime}}{h_G}-\frac{G}{r} u_1^{(2)}+(\e-g \sigma) \frac{h_{G+1}}{h_G} v_1^{(2)}+\frac{g \pi}{1+2 G}\left(u_1^{(2)}+2 \sqrt{G(1+G)} u_2^{(2)}\right)=0 ; \\
	& u_2^{(2) \prime}+u_2^{(2)} \frac{h_G^{\prime}}{h_G}+\frac{G+1}{r} u_2^{(2)}-(\e-g \sigma) \frac{h_{G-1}}{h_G} v_2^{(2)}-\frac{g \pi}{1+2 G}\left(u_2^{(2)}-2 \sqrt{G(1+G)} u_1^{(2)}\right)=0 .
\end{aligned}\right. \label{n-G-II}
\eeq
For the outgoing wave solutions, the boundary conditions at $r\to\infty$ for the matrix elements of $V(r)$ and $U(r)$ are set to be
\begin{gather}
	V(r) \to\left(\begin{array}{ll}
		1 & 0 \\
		0 & 1
	\end{array}\right), \quad
    U(r) \to\left(\begin{array}{cc}
		k/(\e+g \s_v) & 0 \\
		0 & k/(\e+g \s_v)
	\end{array}\right). \label{bdcond}
\end{gather}

In order to derive the phase shift, one needs to construct the scattering matrix $S$. The $2\times 2$ submatrix of the $S$-matrix can be constructed by the matrix solution of either upper part $\mathcal{G}(r)$ or lower part $\mathcal{F}(r)$. In the following we will use the matrix solution $\mathcal{G}(r)$ to make the illustration. The scattering wave function can be written as
\beq
\Psi_{S C}(r)=-\mathcal{G}(r)^*+\mathcal{G}(r) S,
\eeq
where $S$ is the scattering matrix which is unitary. It is required that the scattering solution be regular at the origin which yields
\beq
S=\lim _{r \rightarrow 0} H^{-1}_v(k r) V^{-1}(r) V^{*}(r) H^{*}_v(k r),
\eeq
then the scattering phase shift is obtained as
\beq
\delta(k)=\0{1}{2 i} \operatorname{tr} \ln S=\0{1}{2 i} \lim _{r \rightarrow 0} \operatorname{tr} \ln \left(V^{-1}(r) V^{*}(r)\right). \label{phtv}
\eeq
In a similar way the phase shift could be also evaluated by matrix solution $U(r)$ as
\beq
\delta(k)=\0{1}{2 i} \lim _{r \rightarrow 0} \operatorname{tr} \ln \left(U^{-1}(r) U^{*}(r)\right). \label{phtu}
\eeq
Both evaluations will give identical results of the phase shift $\delta(k)$.

In order to regulate the phase shift, one needs to find the subtraction terms of the phase shift by Born expansion. One can expand the the matrix solution $V(r)$ (or $U(r)$) around the free solution, which form is
\begin{align}
	V(r) ={}& 1+g_s V^{(1,0)}(r)+g_p V^{(0,1)}(r)+g_s^2 V^{(2,0)}(r)+g_p^2 V^{(0,2)}(r) \notag \\
	& + g_s g_p V^{(1,1)}(r)+\cdots, \label{exps_V_gsgp}
\end{align}
where $g_s$ and $g_p$ are the artificial coupling constants for the expansion, which are formally defined by
\beq
\s(r)=\s_v+g_s\tilde{\s}(r), \ \ \ \ \pi(r)=\pi_v+g_p\tilde{\pi}(r).
\eeq
After expanding the first order radial equations to different orders, one can set $g_s=g_p=1$ in the end. In the expansion we introduce $V^{(n_s,n_p)}$, where $n_s$ and $n_p$ label the order in the expansion around $\s_v$ and $\pi_v$, respectively. The specific calculation process for the Born subtraction of the phase shift by the first order differential equations is detailed in Appendix B, and those calculations for the phase shifts and the corresponding Born subtractions by the second order differential equations are presented in Appendix C.

\section{Numerical results and discussions}

\subsection{Phase shifts}
\begin{figure}[htbp]
	\centering
        \hspace{-1.5cm}
	\begin{minipage}{0.9\textwidth}
		\centering
		\includegraphics[width=\textwidth]{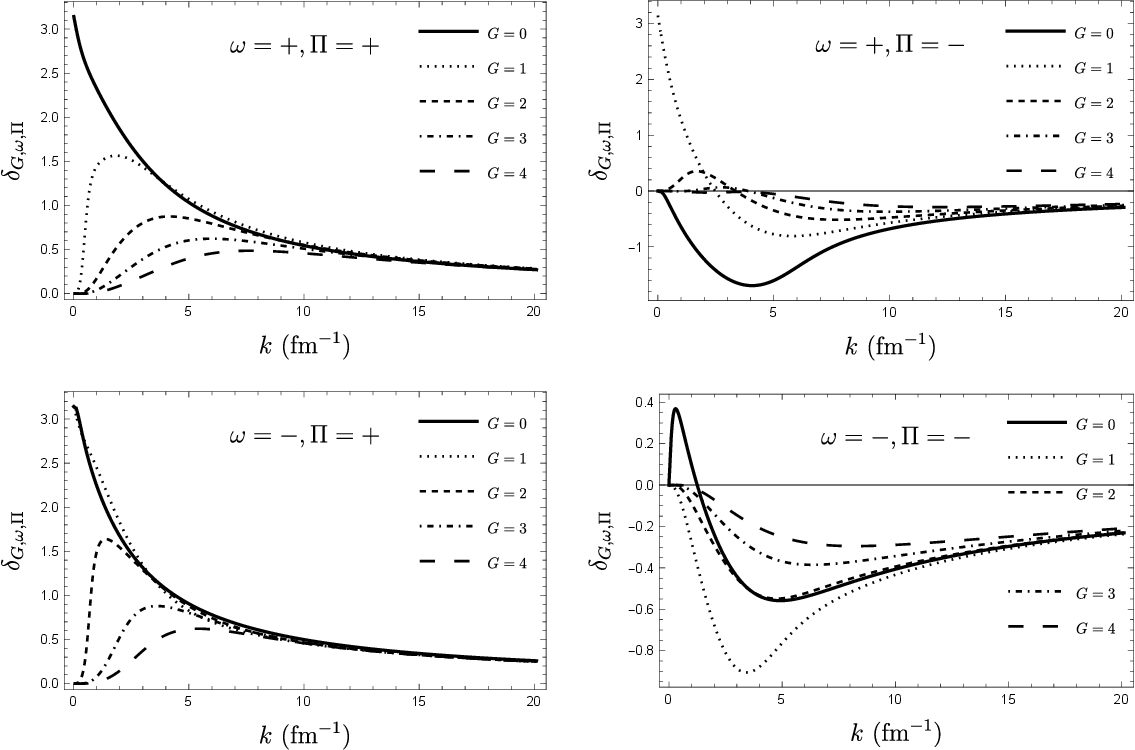}
	\end{minipage}\\
	\caption{The phase shift $\d_{G}$ as a function of momentum $k$ for both parities $\Pi=\pm(-1)^{G}$ and both signs $\omega=\pm$ of the energy.}\label{p-d-G}
\end{figure}

The numerical results of the unsubtracted phase shifts at different grand spin $G$ for both parities $\Pi=\pm(-1)^{G}$ and both signs of the energy $\omega=\pm$ are presented in Fig.\ref{p-d-G}. For the sake of simplicity we use $\Pi=+$ and $\Pi=-$ to represent $\Pi=+(-1)^G$ and $\Pi=-(-1)^G$, respectively. For both parities of $\Pi=+$ and $\Pi=-$ and both signs of the energy $\omega=\pm$, the amplitudes of the phase shifts generally decrease when the value of number $G$ goes large. The absolute values of the phase shift all fall off like $\01k$ with $k\to\infty$, which leads to the logarithmic divergence in the momentum integration. That is why we need to do the phase shift subtraction by the Born expansion. For the parity $\Pi=+$ it could be seen that the values of the phase shifts are all positive. For the case of $G=0$ and $\omega=+$, one can notice that the value of phase shift begins from $\pi$ at $k=0$, which is within expectation. As according to the Levinson theorem \cite{ma_proof_1985}, the difference between the phase shift at the origin and infinity of $k$ is an integer $n$ times of $\pi$,
\beq
\d_G (k=0)-\d_G(k=\infty)=n \pi\label{levinson}
\eeq
where $n$ is the number of bound states. Therefore $n=1$ in $G=0$ case means there is one bound state in this channel, which is exactly the chiral soliton solution in the ground state. For $G=0$ and $G=1$ at $\omega=-$, the phase shifts start from $\pi$ at the threshould ($k=0$), which indicates the existence of one bound state in each of these channels at negative energies.

For the parity $\Pi=-$ and $\omega=+$, the phase shift is negative at $G=0$, while at $G\neq 0$ the value of the phase shift can be either positive or negative, which shows a mild oscillating behavior. There is no bound state for the $G=0$ case which is consistent with our conclusion in the study of the ground state in earlier section. However, for $G=1$ case, the phase shift starts from $\pi$ which indicates there is one positive energy bound state in this case. This bound state could be regarded as a excited state of the chiral soliton. For the parity $\Pi=-$ and $\omega=-$, the phase shift is negative at $G\neq 0$, while at $G=0$ the value of the phase shift can be either positive or negative, which shows a oscillating behavior.

\begin{figure}[htbp]
	\centering
        \hspace{-1.5cm}
	\begin{minipage}{0.9\textwidth}
		\centering
		\includegraphics[width=\textwidth]{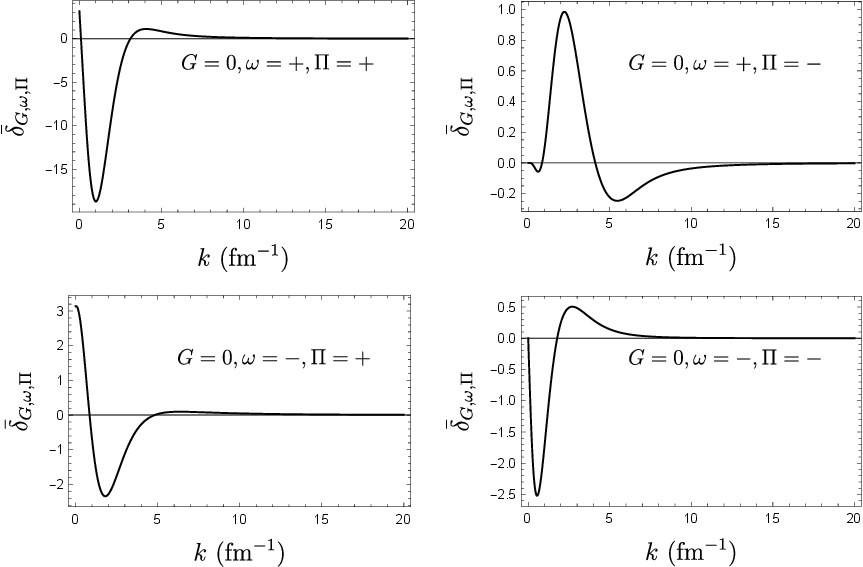}
	\end{minipage}\\
	\caption{The subtracted phase shift function $\bar{\d}_0(k)$ with respect to momentum $k$ in the case of $\Pi=\pm$ and $\omega=\pm$.}\label{G0}
\end{figure}

\begin{figure}[htbp]
	\centering
        \hspace{-1.5cm}
	\begin{minipage}{0.9\textwidth}
		\centering
		\includegraphics[width=\textwidth]{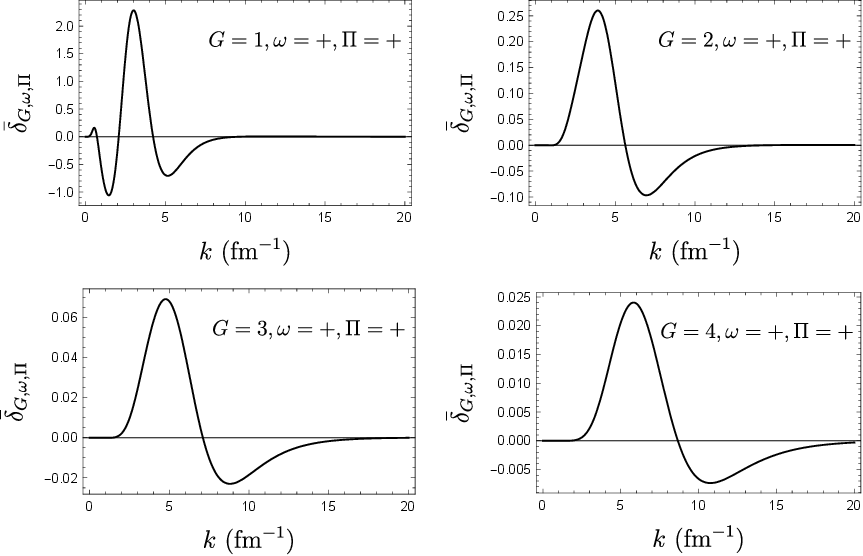}
	\end{minipage}\\
	\caption{The subtracted phase shift $\bar{\d}_G$ as a function of momentum $k$ for $G = 1,\, 2,\, 3,\,4 $ in the case of $\Pi=+$ and $\omega=+$.}\label{ep11}
\end{figure}

\begin{figure}[htbp]
	\centering
        \hspace{-1.5cm}
	\begin{minipage}{0.9\textwidth}
		\centering
		\includegraphics[width=\textwidth]{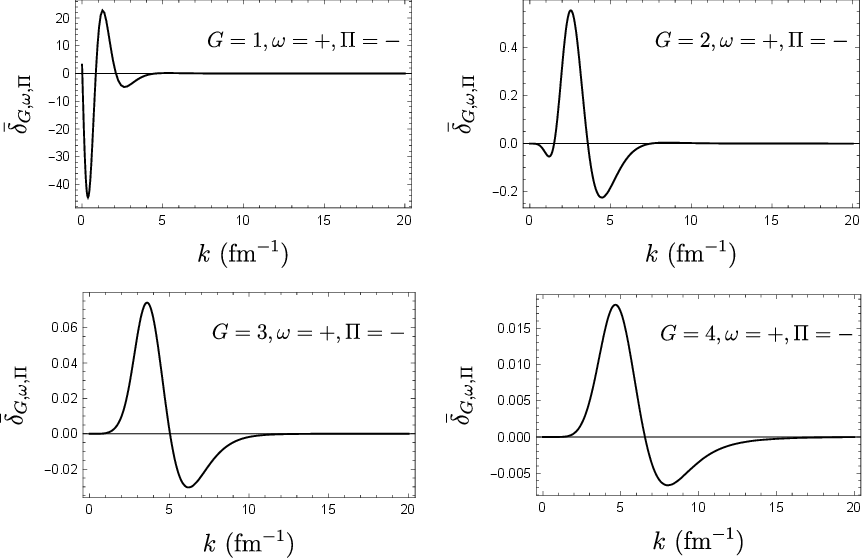}
	\end{minipage}\\
	\caption{The subtracted phase shift $\bar{\d}_G$ as a function of momentum $k$ for $G = 1,\, 2,\, 3,\,4 $ in the case of $\Pi=-$ and $\omega=+$.}\label{ep12}
\end{figure}

\begin{figure}[htbp]
	\centering
        \hspace{-1.5cm}
	\begin{minipage}{0.9\textwidth}
		\centering
		\includegraphics[width=\textwidth]{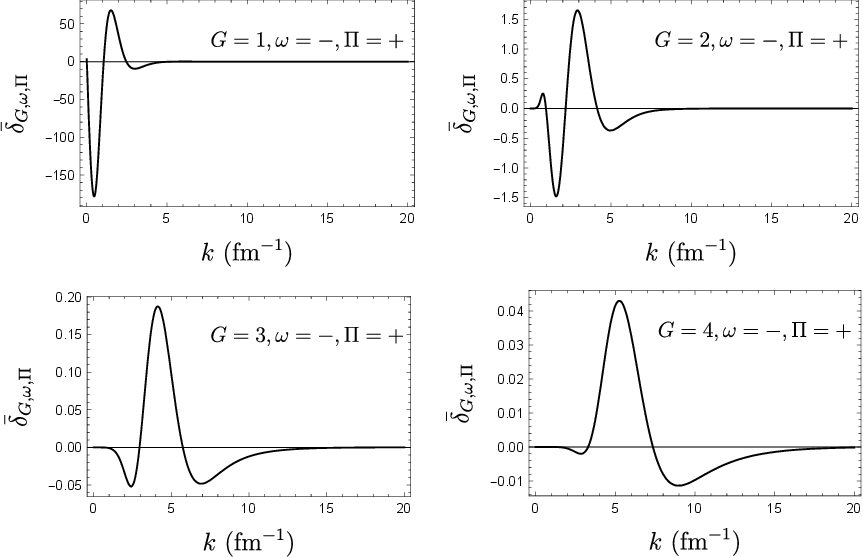}
	\end{minipage}\\
	\caption{The subtracted phase shift $\bar{\d}_G$ as a function of momentum $k$ for $G = 1,\, 2,\, 3,\,4 $ in the case of $\Pi=+$ and $\omega=-$.}\label{ep21}
\end{figure}

\begin{figure}[htbp]
	\centering
        \hspace{-1.5cm}
	\begin{minipage}{0.9\textwidth}
		\centering
		\includegraphics[width=\textwidth]{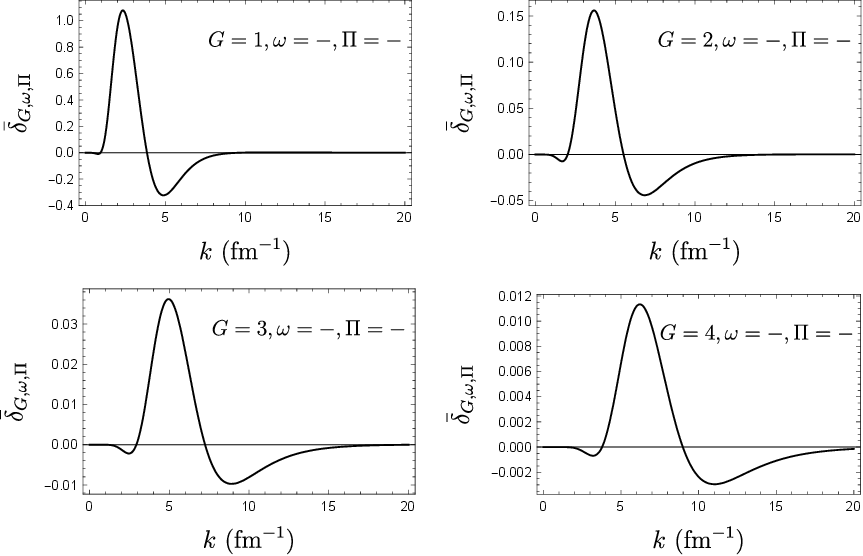}
	\end{minipage}\\
	\caption{The subtracted phase shift $\bar{\d}_G$ as a function of momentum $k$ for $G = 1,\, 2,\, 3,\,4$ in the case of $\Pi=-$ and $\omega=-$.}\label{ep22}
\end{figure}

After the Born subtractions of the phase shifts, the variations of the subtracted phase shifts with momentum $k$ are shown in Fig.\ref{G0}, Fig.\ref{ep11}, Fig.\ref{ep12}, Fig.\ref{ep21} and Fig.\ref{ep22}. One can see that all the phase shifts after subtraction oscillate at small $k$ and approach zero very fast at $k\to\infty$. As a result the divergence in the momentum integration is eliminated. From Fig.\ref{G0}, at $G=0$ in the case of positive energy for the positive parity, the phase shift goes to $0^+$ at $k\to\infty$, while for the negative parity it goes to $0^-$ at $k\to\infty$. The values of the phase shifts for the positive parity at $G=0$ still start from $\pi$, which means the Levinson theorem is not violated after the subtraction. In positive energy case the oscillation amplitude of the phase shift in positive parity is much larger than that in negative parity. In the case of negative energy at $G=0$ for both parities, the phase shifts all go to $0^+$ at $k\to\infty$. The oscillation amplitude of the phase shift in positive parity is also larger than that in negative parity.

Fig.\ref{ep11}, Fig.\ref{ep12}, Fig.\ref{ep21} and Fig.\ref{ep22} give the variations of the subtracted phase shifts at $G\neq 0$ for $\Pi=\pm$ and $\omega=\pm$. It can be seen that all the phase shifts obtained after subtraction converge well to zero at momentum $k\to\infty $. In general the subtract phase shifts exhibit oscillation behavior with the variation of $k$. The general oscillation amplitudes decrease with number $G$ increasing. The subtracted phase shifts at $G=1$ for $\Pi=-$ and $\omega=+$ and for $\Pi=+$ and $\omega=-$ still start from $\pi$, and their oscillation amplitudes are relatively remarkable. It is interesting that for any channel when there is a bound state in the channel, the oscillation amplitude of the subtracted phase shift in that channel becomes prominent. Another interesting feature is that the variation curves of the phase shift tend to expand along the $k$ axis with $G$ increasing.

Finally we should mention that the above phase shifts and subtracted phase shifts are evaluated by the first order differential equations. However, we have also checked that all the phase shifts could be evaluated by the second order differential equations as shown in Appendix C, and it gives exactly the same results as those evaluated by the first order differential equations.

\subsection {Quantum fluctuation energies}
By the subtracted phase shift, the finite quantum fluctuation energy of the chiral soliton at given $G$, $\omega$ and $\Pi$ could be evaluated as
\beq
E_{G,\omega,\Pi}=(2G+1)\int_0^\infty \0{{\rm d}k}{\pi}\0{d\bar\d_{G,\omega,\Pi}(k)}{dk}|E_{q}(k)|. \label{eqEcl_vac_chiral}
\eeq
The whole renormalized quantum fluctuation energy of chiral soliton is expressed as:
\beq
E_{\rm vac}^{\rm ren}[\tilde\s,\tilde\pi]=-\frac{1}{2}\underset{n}{\sum}(2G_{n}+1)|E_{n}|
-\frac{1}{2}\underset{G,\omega,\Pi}{\sum}E_{G,\omega,\Pi}+\Gamma_2+\Gamma_4,\label{ren evac}
\eeq
$G_{n}$ represents the quantum number of grand spin corresponding to the energy level of the discrete bound state $E_{n}$.

The numerical results of the quantum fluctuation energies of the continuum energy spectrum at different $G,\omega$ and $\Pi$ are presented in table \ref{t_E_vac}. For $G=0$ the absolute values or the magnitudes of the quantum fluctuation energies at $\omega=+$, $\Pi=+$ and $\omega=-$, $\Pi=+$ are larger than the others due to the bound state energy levels in those channels. One could also see that for the case of $G=1$, in the channels where there are bound state levels the magnitudes of the quantum fluctuation energies are remarkable. For $G\geq 1$ the magnitudes of the quantum fluctuation energies decrease with $G$ increasing for both parities and energy signs. Due to the $2G+1$ factor, even at certain large $G$ the energy contribution from the partial wave to the entire vacuum polarized energy is still not small.

The numerical results of the discrete bound energy level $E_{n}$ from different quantum numbers ($G$ and $\Pi$) are shown in the Table \ref{t_E_bd}. Among them, $E_{1}=+0.216~\rm fm^{-1}$ represents energy level of the ground state which has already been occupied by the three valence quarks. Due to the Pauli exclusion principle, the sea quarks from vacuum fluctuations can not stay at that energy level. Thus the ground state energy level is not included in the energy level summation of the vacuum energy.
\begin{table}[tbh]
	\centering
	\caption{The quantum fluctuation correction energy $E_{G,\omega,\Pi}$ for $G = 0,\, 1,\, 2,\, 3,\,4 $ at different parities and energy signs.}
	\begin{minipage}[s]{\textwidth}
		\centering
		\[\begin{array}{cccccc}
			\hline
			E_{G,\omega,\Pi}~(\rm fm^{-1}) ~~& G=0 & G=1 & G=2 & G=3 & G=4 \\
			\hline
			\omega=+,\Pi=+ & 0.509 & -0.472 & -0.328 & -0.184 & -0.123 \\
			\omega=+,\Pi=- & -0.136 & -7.785 & -0.190 & -0.082 & -0.044 \\
			\omega=-,\Pi=+ & -1.780 & -6.015 & -0.329 & -0.181 & -0.113 \\
			\omega=-,\Pi=- & -0.114 & -0.498 & -0.187 & -0.085 & -0.052 \\
			\hline
		\end{array} \label{t_E_vac}\]
	\end{minipage}
\end{table}
\begin{table}[tbh]
	\centering
	\caption{The discrete bound energy level $E_{n}$ for $G =0,\, 1 $ at different parities.}
	\begin{minipage}[s]{\textwidth}
		\centering
		\[\begin{array}{cc}
			\hline
			{G,\Pi} & ~~~E_{n}~(\rm fm^{-1})\\
			\hline
			G=0,\Pi=+ & +0.216  \\
			G=0,\Pi=+ & -2.475 \\
			G=1,\Pi=- & +2.430  \\
			G=1,\Pi=+ & -1.868  \\
                \hline
		\end{array} \label{t_E_bd}\]
	\end{minipage}
\end{table}

 In table \ref{Ecl_vac_chiral} we present the different parts of the energy related to the chiral soliton. One can see that after the summation of $G$, $\omega$ and $\Pi$, the quantum fluctuation energy from the continuum energy spectrum is positive and constitutes the largest part in the vacuum energy, while the quantum fluctuation energy from discrete levels is negative and its magnitude is comparable to that from the continuum spectrum. The renormalization energies from the Feynman diagrams are dominated by $\Gamma_2$ which value is negative, while $\Gamma_4$ is positive and relatively small. By adding them all one obtains the total renormalized vacuum fluctuation energy over the chiral soliton background which is negative.
If the color degrees of freedom of quarks are taken into account, the vacuum energy should be
multiplied by a factor of 3. One could see that the quantum fluctuation energy in magnitude is not small compared to the classical energy of the chiral soliton.

\begin{table}[tbh]
	\centering
	\caption{The classical energy and quantum fluctuation energy corrections of the  chiral soliton.}
	\begin{minipage}[s]{\textwidth}
		\centering
		\begin{tabular}{ccccccc}
			\hline
			$E$  \hspace{0.2cm} &  \hspace{0.2cm}
			$E_{\rm cl}$\hspace{0.2cm} &  \hspace{0.2cm}
			$E_{\rm vac}^{\rm ren}$  \hspace{0.2cm} & \hspace{0.2cm}
            $-\frac{1}{2}\underset{n}{\sum}(2G_{n}+1)|E_{n}|$ \hspace{0.2cm} &  \hspace{0.2cm}
            $-\frac{1}{2}\underset{G,\omega,\Pi}{\sum}E_{G,\omega,\Pi}$ \hspace{0.2cm} &  \hspace{0.2cm}
			$\Gamma_2$  \hspace{0.2cm} &  \hspace{0.2cm}
			$\Gamma_4$ \\
			\hline
			$(\rm fm^{-1})$   \hspace{0.2cm} & \hspace{0.2cm}
			+5.827  \hspace{0.2cm} & \hspace{0.2cm}
			-0.789   \hspace{0.2cm} & \hspace{0.2cm}
                -7.684   \hspace{0.2cm} & \hspace{0.2cm}
			+9.097   \hspace{0.2cm} & \hspace{0.2cm}
			-4.471  \hspace{0.2cm} & \hspace{0.2cm}
			+2.269 \\
			\hline
		\end{tabular} \label{Ecl_vac_chiral}
	\end{minipage}
\end{table}

\section{SUMMARY AND OUTLOOK}
In this paper, the quantum fluctuation energies on the chiral soliton background are studied under the framework of the QCD effective model. As the background field is spatially inhomogeneous, the calculation of one loop quantum fluctuation is very nontrivial. The scattering phase shift has been thoroughly analyzed and evaluated, which determines the density of states function of momentum in the loop integration of the energy. The vacuum quantum fluctuation energy on the chiral soliton background has been accurately evaluated. In this work we mainly focus on making the whole calculation scheme of the quantum fluctuation energy over the inhomogeneous field background more efficient and practical. As our final goal is trying to understand non-perturbative QCD, in the future work we hope to develop this calculation scheme to study the energy composition or mass distribution of hadrons and furthermore explore the phase structure of quark matter with a nontrivial QCD vacuum background.

\appendix
\section{Renormalization by the fake boson method}
\begin{figure}[htbp]
	\centering
	\hspace{-1.5cm}
	\begin{minipage}{0.4\textwidth}
		\centering
		\includegraphics[width=\textwidth]{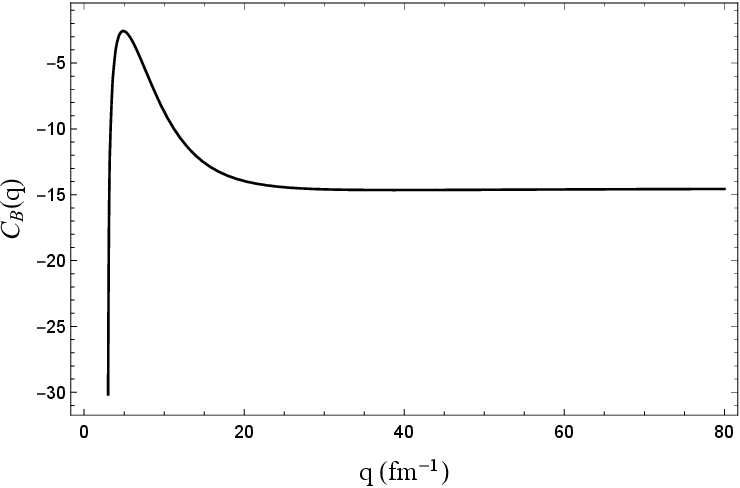}
	\end{minipage}\\
	\caption{The ratio $C_B$ as a function of momentum $q$ at $\rm w = 2$ and $m_{B}=2.5g\sigma_{v}$.}\label{beta}
\end{figure}

In this appendix, we perform renormalization calculations about the logarithmic divergence using the fake boson method. We introduce the fake boson in a background potential
\beq
V_{B}(r)=m_{B}^2 \frac{r}{\rm w} e^{-3\frac{r}{\rm w}},
\eeq
where $\rm w$ is an arbitrary width which should not play a role in the final result, and $m_{B}$ is the mass of the fake boson. We consider the spherically symmetric problem
\beq
-\frac{d^{2}}{d r^{2}} u_{\ell}(r)+\left[\frac{\ell(\ell+1)}{r^{2}}+V_{B}(r)\right] u_{\ell}(r)=k^{2} u_{\ell}(r).
\eeq
In order to obtain the second order scattering phase shift with different angular quantum number $\ell$, we take the ansatz $u_{\ell}(r)=\exp(2i \beta_{\ell}(k,r))rh_{\ell}(kr)$
and the expansion $\beta_{\ell}(k,r)=g_{B} \beta_{\ell}^{(1)}(k,r)+g_{B}^2 \beta_{\ell}^{(2)}(k,r)$, where $g_{B}$ is a formal coupling constant. Then the coupled differential equations about $\beta_{\ell}^{(1)}$
and $\beta_{\ell}^{(2)}$ could be derived as
\begin{align}
-i \beta_{\ell}^{(1)\prime \prime}-2 i k p_{\ell}(k r) \beta_{\ell}^{(1)\prime}+\frac{1}{2} V_{B}(r) & = 0,\\
\\
-i \beta_{\ell}^{(2)\prime \prime}-2 i k p_{\ell}(k r) \beta_{\ell}^{(2)\prime}+2\left(\beta_{\ell}^{(1)\prime}\right)^{2} & = 0 .
\end{align}
By numerically solving the equations, one can obtain the Born subtraction phase shift of the second order as
\beq
\delta_{\ell,B}^{(2)}(k)=2 Re [\beta_{\ell}^{(2)}(k,0)].
\eeq
The logarithmic divergent energies from the second-order fake boson subtraction phase shift and from the third- and fourth-order fermion subtraction phase shifts could be evaluated in the following
\begin{align}
E_{B}^{(2)}(q)&=\int_0^q \0{{\rm d}k}{\pi}\0{d\delta_{B}^{(2)}(k)}{dk}
\sqrt{k^2+m_{B}^2},\\
E_{F}^{(3,4)}(q)&=\int_0^q \0{{\rm d}k}{\pi}\0{d\delta_{F}^{(3,4)}(k)}{dk}
\sqrt{k^2+m_{q}^2},
\end{align}
where the Born phase shifts $\delta_{B}^{(2)}$ and $\delta_{F}^{(3,4)}$ are defined as
\begin{align}
\delta_{B}^{(2)}(k)&=2\sum_{\ell}(2\ell+1)\delta_{\ell,B}^{(2)}(k),
\\
\delta_{F}^{(3,4)}(k)&=\sum_{G,\omega,\Pi}(2G+1)
(\delta_{G,\omega,\Pi}^{(3)}(k)+\delta_{G,\omega,\Pi}^{(4)}(k)).
\end{align}
Next we follow the strategy in Ref.~\cite{cosmicstring} to match the renormalization energies of the fake boson and fermion diagrams. The one loop fake boson diagram related to the second order Born subtraction could be formally evaluated by a on-shell mass renormalization and the result is
\begin{align}
\Pi_{ren}(q)=-\int_{0}^{1}dx\frac{x(1-x)}{m_B^2-x(1-x)m_f^2}(q^2+m_f^2)-\int_{0}^{1}dx\ln\frac{m_B^2+x(1-x)q^2}{m_B^2-x(1-x)m_f^2},
\end{align}
where $m_f$ is a formal mass of the external line of the boson loop. The corresponding finite fake boson renormalization energy is
\begin{align}
\Gamma _{2B}=\int_0^{\infty} \frac{dqq^2}{2\pi ^2}\tilde{V}_{B}(q)\tilde{V}_{B}(-q)\Pi_{ren}(q).
\end{align}
The fake boson renormalization energy and the fermion renormalization energy could be matched by the following scaling constant
\begin{align}
c_{B}=\frac{\Gamma _{4}}{\Gamma_{2B}}, \ \ \ c_{B}\equiv\lim_{q \to \infty}C_{B}(q)=\frac{E_{F}^{(3,4)}|_{q \to \infty}}{E_{B}^{(2)}|_{q \to \infty}},
\end{align}
which can be numerically evaluated by the ratio of $E_{F}^{(3,4)}(q)$ and $E_{B}^{(2)}(q)$ when the momentum ${q \to \infty}$. In the numerical evaluation the width parameter is set to ${\rm w}=2$, then the fake boson mass is determined at $m_{B}=2.5g\sigma_{v}$ to guarantee the scaling parameter approaching a constant when the momentum goes to infinity. The momentum dependent scaling parameter $C_{B}(q)$ is plotted in Fig.\ref{beta}. The numerical value of the scaling constant is fixed at $c_{B}=-14.564$. Thus one obtain the scaled renormalization energy which is equivalent to the renormalization energy of the fermion diagram
\begin{align}
\Gamma _{4}=c_{B} \, \Gamma_{2B}.
\end{align}
However, in this renormalizaiton energy the mass parameter $m_f$ is unphysical and arbitrary. In the main text as we have adopted the renormalization scheme from Farhi's work~\cite{farhi2002searching}, the scaled renormalization energy calculated here should agree with the energy calculated according to the renormalization scheme in Farhi's paper. In the chiral soliton model, by the Farhi's scheme the renormalization energy could be written as
\begin{align}
	\Gamma_{\rm loc}^{(4)}= & \frac{g^4}{8 \pi}\left(\frac{m_{s}^{2}}{m_{q}^2}+6 \int_0^1 d x \ln \left[1-x(1-x) \frac{m_{s}^{2}}{m_{q}^2}\right]\right)  \nnu \\
	& \times \int_0^{\infty} d r r^2\left[\left(\s(r)^2+\pi(r)^2-\s_{v}^2\right)^2-4\s_{v}^{2}(\s(r)-\s_{v})^2\right],
\end{align}
where $m_{s}$ and $m_{q}$ are the physical masses. The subscript denotes that the local approximation has been applied, which means the external momentum in the loop Feynman diagram is set to zero. Accordingly, our scaled renormalization energy at the local approximation could be written as  \begin{align}
\Gamma _{4,\rm loc}=\frac{c_{B}}{4\pi}\left(-\int_{0}^{1}dx\frac{x(1-x)m_f^2}{m_B^2-x(1-x)m_f^2}+\int_{0}^{1}dx\ln\left[1-x(1-x)\frac{m_f^2}{m_B^2}\right] \right)\int_0^{\infty}drr^2 V_B(r)^2.
\end{align}
The two energies, $\Gamma _{4,\rm loc}$ and $\Gamma_{\rm loc}^{(4)}$, must be equal. With the masses $m_{s}$ and $m_{q}$ set to their physical values, the mass parameter $m_f$ is subsequently determined to be $m_f=0.923g\s_v$. This yields a final value for the renormalization energy of $\Gamma_4=2.269{\rm fm}^{-1}$.

\section{Calculations of Born subtractions of phase shifts by the first order differential equations}
In this appendix we will illustrate how to calculate the Born subtraction of the phase shift by the expansions based on the first order differential equations. Although the Born expansion in our case has been carried out to the fourth order, due to the length limit we only present calculations of the Born subtraction to the second order. The evaluations to the fourth order is straightforward but tedious. It has been done in our numerical programm package.

By the Born expansion of matrix solution $V(r)$ and $U(r)$, one can pick out the matrix elements $v_i^{(j)}$ and $u_i^{(j)}$ with the expanded forms and substitute them into the radial equations \eqref{n-G-I} and \eqref{n-G-II}. Then the equations of different orders of the expansion parameters $g_s$ and $g_p$ could be separated out order by order. We can obtain the following equations by the orders of $g_s$ and $g_p$ at two boundary conditions:

\begin{enumerate}
\item $g_s$ order equations:
\beq
	\left\{\begin{aligned}
		&v_1^{(1)(1,0) \prime}+v_1^{(1)(1,0)} \frac{h_{G+1}^{\prime}}{h_{G+1}}+\frac{G+2}{r} v_1^{(1)(1,0)}-\left(\e-g \sigma_v\right) \frac{h_G}{h_{G+1}} u_1^{(1)(1,0)}+\beta \sigma_{s}=0 ;\\
		&v_2^{(1)(1,0) \prime}+v_2^{(1)(1,0)} \frac{h_{G-1}^{\prime}}{h_{G-1}}-\frac{G-1}{r} v_2^{(1)(1,0)}+\left(\e-g \sigma_{v}\right) \frac{h_{G}}{h_{G-1}} u_2^{(1)(1,0)}=0 \text {; }\\
		&u_1^{(1)(1,0) \prime}+u_1^{(1)(1,0)} \frac{h_{G}^{\prime}}{h_{G}}-\frac{G}{r} u_1^{(1)(1,0)}+\left(\e+g \sigma_{v}\right) \frac{h_{G+1}}{h_{G}} v_1^{(1)(1,0)}+\sigma_{s}=0 ;\\
		&u_2^{(1)(1,0) \prime}+u_2^{(1)(1,0)} \frac{h_{G}^{\prime}}{h_{G}}+\frac{G+1}{r} u_2^{(1)(1,0)}-\left(\e+g \sigma_{v}\right) \frac{h_{G-1}}{h_{G}} v_2^{(1)(1,0)}=0 ;
	\end{aligned}\right.
	\eeq
	
	\beq
	\left\{\begin{aligned}
		&v_1^{(2)(1,0) \prime}+v_1^{(2)(1,0)} \frac{h_{G+1}^{\prime}}{h_{G+1}}+\frac{G+2}{r} v_1^{(2)(1,0)}-\left(\e-g \sigma_v\right) \frac{h_G}{h_{G+1}} u_1^{(2)(1,0)}=0 ;\\
		&v_2^{(2)(1,0) \prime}+v_2^{(2)(1,0)} \frac{h_{G-1}^{\prime}}{h_{G-1}}-\frac{G-1}{r} v_2^{(2)(1,0)}+\left(\e-g \sigma_{v}\right) \frac{h_{G}}{h_{G-1}} u_2^{(2)(1,0)}-\beta \sigma_{s}=0 ;\\
		&u_1^{(2)(1,0) \prime}+u_1^{(2)(1,0)} \frac{h_{G}^{\prime}}{h_{G}}-\frac{G}{r} u_1^{(2)(1,0)}+\left(\e+g \sigma_{v}\right) \frac{h_{G+1}}{h_{G}} v_1^{(2)(1,0)}=0 ;\\
		&u_2^{(2)(1,0) \prime}+u_2^{(2)(1,0)} \frac{h_{G}^{\prime}}{h_{G}}+\frac{G+1}{r} u_2^{(2)(1,0)}-\left(\e+g \sigma_{v}\right) \frac{h_{G-1}}{h_{G}} v_2^{(2)(1,0)}-\sigma_{s}=0 ;
	\end{aligned}\right.
	\eeq
	
	\item $g_s^2$ order equations:
	\beq
	\left\{\begin{aligned}
		& v_1^{(1)(2,0)\prime}+v_1^{(1)(2,0)} \frac{h_{G+1}^{\prime}}{h_{G+1}}+\frac{G+2}{r} v_1^{(1)(2,0)}-\left(\e-g \sigma_{v}\right) \frac{h_{G}}{h_{G+1}} u_1^{(1)(2,0)}+\sigma_{s} u_1^{(1)(1,0)}=0 ; \\
		& v_2^{(1)(2,0)\prime}+v_2^{(1)(2,0)} \frac{h_{G-1}^{\prime}}{h_{G-1}}-\frac{G-1}{r} v_2^{(1)(2,0)}+\left(\e-g \sigma_v\right) \frac{h_G}{h_{G-1}} u_2^{(1)(2,0)}-\sigma_s u_2^{(1)(1,0)}=0 ; \\
		& u_1^{(1)(2,0)\prime}+u_1^{(1)(2,0)} \frac{h_{G}^{\prime}}{h_{G}}-\frac{G}{r}u_1^{(1)(2,0)}+\left(\e+g \sigma_{v}\right) \frac{h_{G+1}}{h_{G}} v_1^{(1)(2,0)}+\sigma_{s} v_1^{(1)(1,0)}=0 ; \\
		& u_2^{(1)(2,0)\prime}+u_2^{(1)(2,0)} \frac{h_G^{\prime}}{h_G}+\frac{G+1}{r} u_2^{(1)(2,0)}-\left(\e+g \sigma_v\right) \frac{h_{G-1}}{h_G} v_2^{(1)(2,0)}-\sigma_s v_2^{(1)(1,0)}=0 ;
	\end{aligned}\right.
	\eeq
	
	\beq
	\left\{\begin{aligned}
		& v_1^{(2)(2,0)\prime}+v_1^{(2)(2,0)} \frac{h_{G+1}^{\prime}}{h_{G+1}}+\frac{G+2}{r} v_1^{(2)(2,0)}-\left(\e-g \sigma_{v}\right) \frac{h_{G}}{h_{G+1}} u_1^{(2)(2,0)}+\sigma_{s} u_1^{(2)(1,0)}=0 ; \\
		& v_2^{(2)(2,0)\prime}+v_2^{(2)(2,0)} \frac{h_{G-1}^{\prime}}{h_{G-1}}-\frac{G-1}{r} v_2^{(2)(2,0)}+\left(\e-g \sigma_{v}\right) \frac{h_{G}}{h_{G-1}} u_2^{(2)(2,0)}-\sigma_{s} u_2^{(2)(1,0)}=0 ; \\
		& u_1^{(2)(2,0)\prime}+u_1^{(2)(2,0)} \frac{h_{G}^{\prime}}{h_{G}}-\frac{G}{r} u_1^{(2)(2,0)}+\left(\e+g \sigma_{v}\right) \frac{h_{G+1}}{h_{G}} v_1^{(2)(2,0)}+\sigma_{s} v_1^{(2)(1,0)}=0 ; \\
		& u_2^{(2)(2,0)\prime}+u_2^{(2)(2,0)} \frac{h_{G}^{\prime}}{h_{G}}+\frac{G+1}{r} u_2^{(2)(2,0)}-\left(\e+g \sigma_{v}\right) \frac{h_{G-1}}{h_{G}} v_2^{(2)(2,0)}-\sigma_{s} v_2^{(2)(1,0)}=0 ;
	\end{aligned}\right.
	\eeq
	
	\item $g_p$ order equations:
	
	\beq
	\left\{\begin{aligned}
		& v_1^{(1)(0,1) \prime}+v_1^{(1)(0,1)} \frac{h_{G+1}^{\prime}}{h_{G+1}}+\frac{G+2}{r} v_1^{(1)(0,1)}-\left(\e-g \sigma_v\right) \frac{h_G}{h_{G+1}} u_1^{(1)(0,1)}+\frac{\pi}{1+2 G}=0 ; \\
		& v_2^{(1)(0,1) \prime}+v_2^{(1)(0,1)} \frac{h_{G-1}^{\prime}}{h_{G-1}}-\frac{G-1}{r} v_2^{(1)(0,1)}+\left(\e-g \sigma_v\right) \frac{h_G}{h_{G-1}} u_2^{(1)(0,1)}+\frac{2 \sqrt{G(1+G)} \pi}{1+2 G}=0 \text {; } \\
		& u_1^{(1)(0,1) \prime}+u_1^{(1)(0, 1)} \frac{h_G^{\prime}}{h_G}-\frac{G}{r} u_1^{(1)(0, 1)}+\left(\e+g \sigma_v\right) \frac{h_{G+1}}{h_G} v_1^{(1)(0, 1)}+\frac{\beta \pi}{1+2 G}=0 ; \\
		& u_2^{(1)(0,1)\prime}+u_2^{(1)(0,1)} \frac{h_G^{\prime}}{h_G}+\frac{G+1}{r} u_2^{(1)(0, 1)}-\left(\e+g \sigma_v\right) \frac{h_{G-1}}{h_G} v_2^{(1)(0,1)}+\frac{2 \sqrt{G(1+G)} \beta \pi}{1+2 G}=0 ;
	\end{aligned}\right.
	\eeq
	
	\beq
	\left\{\begin{aligned}
		& v_1^{(2)(0,1)\prime}+v_1^{(2)(0,1)} \frac{h^{\prime}{ }_{G+1}}{h_{G+1}}+\frac{G+2}{r} v_1^{(2)(0,1)}-\left(\e-g \sigma_v\right) \frac{h_G}{h_{G+1}} u_1^{(2)(0,1)}+\frac{2 \sqrt{G(1+G)} \pi}{1+2 G}=0 ;\\	
		& v_2^{(2)(0,1) \prime}+v_2^{(2)(0, 1)} \frac{h_{G-1}^{\prime}}{h_{G-1}}-\frac{G-1}{r} v_2^{(2)(0, 1)}+\left(\e-g \sigma_v\right) \frac{h_G}{h_{G-1}} u_2^{(2)(0, 1)}+\frac{\pi}{1+2 G}=0 ; \\
		& u_1^{(2)(0,1) \prime}+u_1^{(2)(0, 1)} \frac{h_G^{\prime}}{h_G}-\frac{G}{r} u_1^{(2)(0, 1)}+\left(\e+g \sigma_v\right) \frac{h_{G+1}}{h_G} v_1^{(2)(0,1)}+\frac{2 \sqrt{G(1+G)} \beta \pi}{1+2 G}=0 \text {; } \\
		& u_2^{(2)(0,1) \prime}+u_2^{(2)(0,1)} \frac{h_G^{\prime}}{h_G}+\frac{G+1}{r} u_2^{(2)(0, 1)}-\left(\e+g \sigma_v\right) \frac{h_{G-1}}{h_G} v_2^{(2)(0,1)}-\frac{\beta \pi}{1+2 G}=0 ;
	\end{aligned}\right.
	\eeq	
	
	\item $g_p^2$ order equations:
	
	\beq
	\hspace{-1cm}
	\left\{\begin{aligned}
		& v_1^{(1)(0,2) \prime}+v_1^{(1)(0,2)} \frac{h_{G+1}^{\prime}}{h_{G+1}}+\frac{G+2}{r} v_1^{(1)(0,2)}-\left(\e-g \sigma_v\right) \frac{h_G}{h_{G+1}} u_1^{(1)(0,2)}+\frac{\pi}{1+2 G} v_1^{(1)(0,1)}-\frac{2 \sqrt{G(1+G)} \pi}{1+2 G} v_2^{(1)(0,1)}=0 ; \\
		& v_2^{(1)(0,2) \prime}+v_2^{(1)(0,2)} \frac{h_{G-1}^{\prime}}{h_{G-1}}-\frac{G-1}{r} v_2^{(1)(0,2)}+\left(\e-g \sigma_v\right) \frac{h_G}{h_{G-1}} u_2^{(1)(0,2)}+\frac{\pi}{1+2 G} v_2^{(1)(0,1)}+\frac{2 \sqrt{G(1+G)} \pi}{1+2 G} v_1^{(1)(0,1)}=0 ; \\
		& u_1^{(1)(0,2)\prime}+u_1^{(1)(0,2)} \frac{h_G^{\prime}}{h_G}-\frac{G}{r} u_1^{(1)(0,2)}+\left(\e+g \sigma_v\right) \frac{h_{G+1}}{h_G} v_1^{(1)(0,2)}+\frac{\pi}{1+2 G} u_1^{(1)(0,1)}+\frac{2 \sqrt{G(1+G)} \pi}{1+2 G} u_2^{(1)(0,1)}=0 ; \\
		& u_2^{(1)(0,2)\prime}+u_2^{(1)(0,2)} \frac{h_G^{\prime}}{h_G}+\frac{G+1}{r} u_2^{(1)(0,2)}-\left(\e+g \sigma_v\right) \frac{h_{G-1}}{h_G} v_2^{(1)(0,2)}-\frac{\pi}{1+2 G} u_2^{(1)(0,1)}+\frac{2 \sqrt{G(1+G)} \pi}{1+2 G} u_1^{(1)(0,1)}=0 ;
	\end{aligned}\right.
	\eeq
	
	\beq
	\hspace{-1cm}
	\left\{\begin{aligned}
		& v_1^{(2)(0,2) \prime}+v_1^{(2)(0,2)} \frac{h_{G+1}^{\prime}}{h_{G+1}}+\frac{G+2}{r} v_1^{(2)(0,2)}-\left(\e-g \sigma_v\right) \frac{h_G}{h_{G+1}} u_1^{(2)(0,2)}+\frac{\pi}{1+2 G} v_1^{(2)(0,1)}-\frac{2 \sqrt{G(1+G)} \pi}{1+2 G} v_2^{(2)(0,1)}=0 ; \\
		& v_2^{(2)(0,2) \prime}+v_2^{(2)(0,2)} \frac{h_{G-1}^{\prime}}{h_{G-1}}-\frac{G-1}{r} v_2^{(2)(0,2)}+\left(\e-g \sigma_v\right) \frac{h_G}{h_{G-1}} u_2^{(2)(0,2)}+\frac{\pi}{1+2 G} v_2^{(2)(0, 1)}+\frac{2 \sqrt{G(1+G)} \pi}{1+2 G} v_1^{(2)(0, 1)}=0 ; \\
		& u_1^{(2)(0,2) \prime}+u_1^{(2)(0,2)} \frac{h_G^{\prime}}{h_G}-\frac{G}{r} u_1^{(2)(0,2)}+\left(\e+g \sigma_v\right) \frac{h_{G+1}}{h_G} v_1^{(2)(0, 2)}+\frac{\pi}{1+2 G} u_1^{(2)(0, 1)}+\frac{2 \sqrt{G(1+G)} \pi}{1+2 G} u_2^{(2)(0, 1)}=0 ; \\
		& u_2^{(2)(0,2)\prime}+u_2^{(2)(0,2)} \frac{h_G^{\prime}}{h_G}+\frac{G+1}{r} u_2^{(2)(0,2)}-\left(\e+g \sigma_v\right) \frac{h_{G-1}}{h_G} v_2^{(2)(0,2)}-\frac{\pi}{1+2 G} u_2^{(2)(0,1)}+\frac{2 \sqrt{G(1+G)} \pi}{1+2 G} u_1^{(2)(0,1)}=0 ;
	\end{aligned}\right.
	\eeq
	
	\item $g_s g_p$ order equations:
	
	\beq
	\hspace{-2.7cm}
	\left\{\begin{aligned}
		& v_1^{(1)(1,1) \prime}+v_1^{(1)(1,1)} \frac{h_{G+1}^{\prime}}{h_{G+1}}+\frac{G+2}{r} v_1^{(1)(1,1)}-\left(\e-g \sigma_v\right) \frac{h_G}{h_{G+1}} u_1^{(1)(1,1)}+\sigma_s u_1^{(1)(0,1)}+\frac{\pi}{1+2 G} v_1^{(1)(1,0)}-\frac{2 \sqrt{G(1+G) \pi}}{1+2 G} v_2^{(1)(1,0)}=0 ; \\
		& v_2^{(1)(1,1) \prime}+v_2^{(1)(1,1)} \frac{h_{G-1}^{\prime}}{h_{G-1}}-\frac{G-1}{r} v_2^{(1)(1,1)}+\left(\e-g \sigma_v\right) \frac{h_G}{h_{G-1}} u_2^{(1)(1,1)}-\sigma_s u_2^{(1)(0,1)}+\frac{\pi}{1+2 G} v_2^{(1)(1,0)}+\frac{2 \sqrt{G(1+G)} \pi}{1+2 G} v_1^{(1)(1,0)}=0 ; \\
		& u_1^{(1)(1,1)\prime}+u_1^{(1)(1,1)} \frac{h_G^{\prime}}{h_G}-\frac{G}{r} u_1^{(1)(1,1)}+\left(\e+g \sigma_v\right) \frac{h_{G+1}}{h_G} v_1^{(1)(1,1)}+\sigma_s v_1^{(1)(0,1)}+\frac{\pi}{1+2 G} u_1^{(1)(1,0)}+\frac{2 \sqrt{G(1+G)} \pi}{1+2 G} u_2^{(1)(1,0)}=0 ; \\
		& u_2^{(1)(1,1)\prime}+u_2^{(1)(1,1)} \frac{h_G^{\prime}}{h_G}+\frac{G+1}{r} u_2^{(1)(1,1)}-\left(\e+g \sigma_v\right) \frac{h_{G-1}}{h_G} v_2^{(1)(1,1)}-\sigma_s v_2^{(1)(0,1)}-\frac{\pi}{1+2 G} u_2^{(1)(1,0)}+\frac{2 \sqrt{G(1+G)} \pi}{1+2 G} u_1^{(1)(1,0)}=0 ;
	\end{aligned}\right.
	\eeq
	
	\beq
	\hspace{-2.7cm}
	\left\{\begin{aligned}
		& v_1^{(2)(1,1) \prime}+v_1^{(2)(1,1)} \frac{h_{G+1}^{\prime}}{h_{G+1}}+\frac{G+2}{r} v_1^{(2)(1,1)}-\left(\e-g \sigma_v\right) \frac{h_G}{h_{G+1}} u_1^{(2)(1,1)}+\sigma_s u_1^{(2)(0,1)}+\frac{\pi}{1+2 G} v_1^{(2)(1,0)}-\frac{2 \sqrt{G(1+G)} \pi}{1+2 G} v_2^{(2)(1,0)}=0 ; \\
		& v_2^{(2)(1,1) \prime}+v_2^{(2)(1,1)} \frac{h_{G-1}^{\prime}}{h_{G-1}}-\frac{G-1}{r} v_2^{(2)(1,1)}+\left(\e-g \sigma_v\right) \frac{h_G}{h_{G-1}} u_2^{(2)(1,1)}-\sigma_s u_2^{(2)(0,1)}+\frac{\pi}{1+2 G} v_2^{(2)(1,0)}+\frac{2 \sqrt{G(1+G)} \pi}{1+2 G} v_1^{(2)(1,0)}=0 ; \\
		& u_1^{(2)(1,1) \prime}+u_1^{(2)(1,1)} \frac{h_G^{\prime}}{h_G}-\frac{G}{r} u_1^{(2)(1,1)}+\left(\e+g \sigma_v\right) \frac{h_{G+1}}{h_G} v_1^{(2)(1,1)}+\sigma_s v_1^{(2)(0,1)}+\frac{\pi}{1+2 G} u_1^{(2)(1,0)}+\frac{2 \sqrt{G(1+G)} \pi}{1+2 G} u_2^{(2)(1,0)}=0 ; \\
		& u_2^{(2)(1,1) \prime}+u_2^{(2)(1,1)} \frac{h_G^{\prime}}{h_G}+\frac{G+1}{r} u_2^{(2)(1,1)}-\left(\e+g \sigma_v\right) \frac{h_{G-1}}{h_G} v_2^{(2)(1,1)}-\sigma_s v_2^{(2)(0,1)}-\frac{\pi}{1+2 G} u_2^{(2)(1,0)}+\frac{2 \sqrt{G(1+G)} \pi}{1+2 G} u_1^{(2)(1,0)}=0 ;
	\end{aligned}\right.
	\eeq
	
\end{enumerate}

Where $\beta=k /\left(\varepsilon+g \sigma_{v}\right),\, \sigma_{s}=\tilde{\sigma}(r) \text { and } \pi=\tilde{\pi}(r)$. The above equations could be numerically solved order by order at given $k$ and $G$ with parity $\Pi=-(-1)^G$. The equations with parity $\Pi=+(-1)^G$ are similar. The elements $v_i^{(j)(n_s,\,n_p)}$ of different orders can be collected to construct following matrix
\beq
V^{(n_s,\,n_p)}_G\equiv
\begin{pmatrix}
	v_1^{(1)(n_s,\,n_p)} & v_1^{(2)(n_s,\,n_p)} \\
	v_2^{(1)(n_s,\,n_p)} & v_2^{(2)(n_s,\,n_p)}
\end{pmatrix},
\eeq
and we assume that
\begin{align}
	V^{(1)}_G(r) &= V^{(1,0)}_G(r)+V^{(0,1)}_G(r), \\
	V^{(2)}_G(r) &= V^{(2,0)}_G(r)+V^{(0,2)}_G(r)+V^{(1,1)}_G(r).
\end{align}
Then the first and second order Born subtraction terms of the phase shift are defined as
\begin{align}
	\d_G^{(1)}(k) &= \0{1}{2i}\lim_{r\to 0}\operatorname{tr}\left[V_G^{(1)*}(r)-V_G^{(1)}(r)\right], \\
	\d_G^{(2)}(k) &= \0{1}{2i}\lim_{r\to 0}\operatorname{tr}\left[V_G^{(2)*}(r)-V_G^{(2)}(r)-\0{1}{2}\left[V_G^{(1)}(r)\right]^2+\0{1}{2}\left[V_G^{(1)*}(r)\right]^2\right].
\end{align}
The Born subtraction terms could also be calculated by the matrix $U^{(1)}_G(r)$ and $U^{(2)}_G(r)$ with the matrix elements $u_i^{(j)(n_s,\,n_p)}$ in a similar way.

Finally if $G=0$, the Born subtraction terms of the phase shifts are simplified to the forms as
\begin{align}
	\d_0^{(1)}(k) &= \0{1}{2i}\lim_{r\to 0}\operatorname{tr}\left[v^*(r)-v(r)\right], \\
	\d_0^{(2)}(k) &= \0{1}{2i}\lim_{r\to 0}\operatorname{tr}\left[v^*(r)-v(r)-\0{1}{2}v(r)^2+\0{1}{2}v^*(r)^2\right].
\end{align}

\section{Calculations of Born subtractions of phase shifts by the second order differential equations}
In this appendix we will illustrate how to calculate the Born series of the phase shifts by using the second order differential equations. Though it is tedious, the Born expansion in our calculation has been carried out to the fourth order, which has been done in our numerical program package. Due to the length limit we only present expanded equations of the Born expansion to the second order.

From equation \eqref{n-G}, one can decouple the first order coupled equations with 4 functions into the following second order coupled equations with 2 functions. Here we choose to eliminate $f_1$ and $f_2$ to obtain the decoupled equations with functions $g_1$ and $g_2$,
\begin{equation}
	\left\{ \begin{array}{l}
		g_1''+\left( \frac{2}{r}-\frac{\varepsilon _1'}{\varepsilon _1} \right) g_1'+\left[ -\frac{2\sqrt{G\left( G+1 \right)}}{2G+1}\frac{g\pi}{r}+\frac{2\sqrt{G\left( G+1 \right)}}{2G+1}g\pi '-\frac{\varepsilon _1'}{\varepsilon _1}\frac{2\sqrt{G\left( G+1 \right)}}{2G+1}g\pi \right] g_2\\
		
		+\left[ -\frac{\left( G+1 \right) \left( G+2 \right)}{r^2}+\frac{2\left(G+1 \right)}{2G+1}\frac{g\pi}{r}-\frac{\varepsilon _1'}{\varepsilon _1}\frac{G+2}{r}+\frac{\varepsilon _1'}{\varepsilon _1}\frac{g\pi}{2G+1}-\frac{g\pi '}{2G+1}+\varepsilon^{2}-g^2(\sigma^2+\pi^2) \right]g_1=0, \\
		
		g_2''+\left( \frac{2}{r}-\frac{\varepsilon _1'}{\varepsilon _1} \right) g_2'+\left[-\frac{2\sqrt{G\left( G+1 \right)}}{2G+1}\frac{g\pi}{r}+\frac{2\sqrt{G\left( G+1 \right)}}{2G+1}g\pi '-\frac{\varepsilon _1'}{\varepsilon _1}\frac{2\sqrt{G\left( G+1 \right)}}{2G+1}g\pi \right] g_1\\
		
		+\left[-\frac{G\left( G-1 \right)}{r^2}+\frac{2G}{2G+1}\frac{g\pi}{r}+\frac{\varepsilon _1'}{\varepsilon _1}\frac{G-1}{r}-\frac{\varepsilon _1'}{\varepsilon _1}\frac{g\pi}{2G+1}+\frac{g\pi '}{2G+1}+\varepsilon^{2}-g^2(\sigma^2+\pi^2) \right] g_2=0, \\
	\end{array} \right. \label{2nd eqs1}
\end{equation}
where $\varepsilon _1 =\varepsilon+g\sigma$. Substituting the matrix solutions \eqref{vh} into the second order equations \eqref{2nd eqs1}, one can obtain the following differential equations of matrix elements of $V(r)$ for the two boundary conditions separately, which are
\begin{equation}
\left\{\begin{aligned}
	v_{1}^{(1) \prime \prime} & +\frac{2}{r} v_{1}^{(1) \prime}\left(1+\frac{r h_{G+1}^{\prime}}{h_{G+1}}\right)-\frac{\varepsilon_{1}^{\prime}}{\varepsilon_{1}}\left(v_{1}^{(1) \prime}+v_{1}^{(1)} \frac{h_{G+1}^{\prime}}{h_{G+1}}\right)+w_{11} v_{1}^{(1)}+w_{12} v_{2}^{(1)}=0, \\
	v_{2}^{(1) \prime \prime} & +\frac{2}{r} v_{2}^{(1) \prime}\left(1+\frac{r h_{G+1}^{\prime}}{h_{G+1}}\right)-\frac{\varepsilon_{1}^{\prime}}{\varepsilon_{1}}\left(v_{2}^{(1) \prime}+v_{2}^{(1)} \frac{h_{G+1}^{\prime}}{h_{G+1}}\right)+w_{21} v_{1}^{(1)}+w_{22} v_{2}^{(1)} \\
	& +\frac{2(2 G+1)}{r^{2}} v_{2}^{(1)}=0,
\end{aligned}\right. \label{2nd eqs2}
\end{equation}
and
\begin{equation}
\left\{\begin{aligned}
	v_{1}^{(2) \prime \prime} & +\frac{2}{r} v_{1}^{(2) \prime}\left(1+\frac{r h_{G-1}^{\prime}}{h_{G-1}}\right)-\frac{\varepsilon_{1}^{\prime}}{\varepsilon_{1}}\left(v_{1}^{(2) \prime}+v_{1}^{(2)} \frac{h_{G-1}^{\prime}}{h_{G-1}}\right)+w_{11} v_{1}^{(2)}+w_{12} v_{2}^{(2)} \\
	& -\frac{2(2 G+1)}{r^{2}} v_{1}^{(2)}=0, \\
	v_{2}^{(2) \prime \prime} & +\frac{2}{r} v_{2}^{(2) \prime}\left(1+\frac{r h_{G-1}^{\prime}}{h_{G-1}}\right)-\frac{\varepsilon_{1}^{\prime}}{\varepsilon_{1}}\left(v_{2}^{(2) \prime}+v_{2}^{(2)} \frac{h_{G-1}^{\prime}}{h_{G-1}}\right)+w_{21} v_{1}^{(2)}+w_{22} v_{2}^{(2)}=0,
\end{aligned}\right. \label{2nd eqs3}
\end{equation}
and
\begin{equation}
\begin{aligned}
	w_{11}= & -g^{2}\left(\sigma^{2}+\pi^{2}-\sigma_{v}^{2}\right)-\frac{G+2}{r} \frac{\sigma^{\prime}}{\sigma+\varepsilon / g}-\frac{g \pi^{\prime}}{2 G+1} \\
	& +\frac{2}{r} \frac{G+1}{2 G+1} g \pi+\frac{g \pi}{2 G+1} \frac{\sigma^{\prime}}{\sigma+\varepsilon / g}, \\
	w_{22}= & -g^{2}\left(\sigma^{2}+\pi^{2}-\sigma_{v}^{2}\right)+\frac{G-1}{r} \frac{\sigma^{\prime}}{\sigma+\varepsilon / g}+\frac{g \pi^{\prime}}{2 G+1} \\
	& +\frac{2}{r} \frac{G}{2 G+1} g \pi-\frac{g \pi}{2 G+1} \frac{\sigma^{\prime}}{\sigma+\varepsilon / g}, \\
	w_{12}= & w_{21}=\frac{2 \sqrt{G(G+1)}}{2 G+1}\left[g \pi^{\prime}-g \pi\left(\frac{1}{r}+\frac{\sigma^{\prime}}{\sigma+\varepsilon / g}\right)\right] .
\end{aligned}
\end{equation}
By the Born expansion of matrix solution $V(r)$, one can pick out the matrix elements $v_i^{(j)}$ with the expanded forms and substitute them into the second order differential equations \eqref{2nd eqs2} and \eqref{2nd eqs3}. Then the equations of different orders of the expansion parameters $g_s$ and $g_p$ could be separated out order by order. One can obtain the following equations by the orders of $g_s$ and $g_p$ at two boundary conditions:
\begin{enumerate}
\item $g_s$ order equations:
 \hspace{-5cm}
\begin{align}
&\left\{\begin{aligned}
& v_1^{(1)(1,0)\pr\pr}+\0{2}{r}v_1^{(1)(1,0)\pr}\left(1+\0{r h_{G+1}^\pr}{h_{G+1}}\right)+\left(-\0{\s_{s}'}{\e_v}\right)\0{h_{G+1}'}{h_{G+1}} \\
& \qquad +\left(-\0{\s_{s}'}{\e_v}\right)\0{G+2}{r}=0, \\
& v_2^{(1)(1,0)\pr\pr}+\0{2}{r}v_2^{(1)(1,0)}{}'\left(1+\0{r h_{G+1}'}{h_{G+1}}\right)+\0{2(2G+1)}{r^2}v_2^{(1)(1,0)}=0,
\end{aligned}\right. \\
&\left\{\begin{aligned}
& v_1^{(2)(1,0)\pr\pr}+\0{2}{r}v_1^{(2)(1,0)\pr}\left(1+\0{r h_{G-1}^\pr}{h_{G-1}}\right)-\0{2(2G+1)}{r^2}v_1^{(2)(1,0)}=0, \\
& v_2^{(2)(1,0)\pr\pr}+\0{2}{r}v_2^{(2)(1,0)\pr}\left(1+\0{r h_{G-1}^\pr}{h_{G-1}}\right)+\left(-\0{\s_{s}'}{\e_v}\right)\0{h_{G-1}'}{h_{G-1}}+\0{\s_{s}'}{\e_v}\0{G-1}{r}=0,
\end{aligned}\right.
\end{align}

\item $g_s^2$ order equations:
\hspace{-5cm}
\begin{align}
&\left\{\begin{aligned}
& -\0{\s_{s}'}{\e_v}\left(v_1^{(1)(1,0)\pr}+v_1^{(1)(1,0)}\0{h_{G+1}'}{h_{G+1}}\right)+\left(-\0{\s_{s}'}{\e_v}\right)\0{G+2}{r}v_1^{(1)(1,0)}+\0{\s_{s}'\s_{s}}{\e_v^2}\0{h_{G+1}'}{h_{G+1}} \\
& \qquad +\0{\s_{s}'\s_{s}}{\e_v^2}\0{G+2}{r}-(\s_{s}^2+2\s_{v}\s_{s})+v_1^{(1)(2,0)\pr\pr} \\
& \qquad +\0{2}{r}v_1^{(1)(2,0)\pr}\left(1+\0{r h_{G+1}'}{h_{G+1}}\right)=0, \\
& -\0{\s_{s}'}{\e_v}\left(v_2^{(1)(1,0)\pr}+v_2^{(1)(1,0)}\0{h_{G+1}'}{h_{G+1}}\right)+\left(\0{\s_{s}'}{\e_v}\right)\0{G-1}{r}v_2^{(1)(1,0)}+v_2^{(1)(2,0)\pr\pr} \\
& \qquad +\0{2}{r}v_2^{(1)(2,0)\pr}\left(1+\0{r h_{G+1}'}{h_{G+1}}\right)+\0{2(2G+1)}{r^2}v_2^{(1)(2,0)}=0,
\end{aligned}\right. \\
&\left\{\begin{aligned}
& -\0{\s_{s}'}{\e_v}\left(v_1^{(2)(1,0)\pr}+v_1^{(2)(1,0)}\0{h_{G-1}'}{h_{G-1}}\right)+\left(-\0{\s_{s}'}{\e_v}\right)\0{G+2}{r}v_1^{(2)(1,0)}+v_1^{(2)(2,0)\pr\pr} \\
& \qquad +\0{2}{r}v_1^{(2)(2,0)\pr}\left(1+\0{r h_{G-1}'}{h_{G-1}}\right)-\0{2(2G+1)}{r^2}v_1^{(2)(2,0)}=0, \\
& -\0{\s_{s}'}{\e_v}\left(v_2^{(2)(1,0)\pr}+v_2^{(2)(1,0)}\0{h_{G-1}'}{h_{G-1}}\right)+\0{\s_{s}'}{\e_v}\0{G-1}{r}v_2^{(2)(1,0)}+\0{\s_{s}'\s_{s}}{\e_v^2}\0{h_{G-1}'}{h_{G-1}} \\
& \qquad -\0{\s_{s}'\s_{s}}{\e_v^2}\0{G-1}{r}-(\s_{s}^2+2\s_{v}\s_{s})+v_2^{(2)(2,0)\pr\pr} \\
& \qquad +\0{2}{r}v_2^{(2)(2,0)\pr}\left(1+\0{r h_{G-1}'}{h_{G-1}}\right)=0,
\end{aligned}\right.
\end{align}

\item $g_p$ order equations:
\hspace{-5cm}
\begin{align}
&\left\{\begin{aligned}
& v_1^{(1)(0,1)\pr\pr}+\0{2}{r}v_1^{(1)(0,1)\pr}\left(1+\0{r h_{G+1}^\pr}{h_{G+1}}\right)+\0{G+1}{2G+1}\0{2}{r}\pi-\0{\pi '}{2G+1}=0, \\
& v_2^{(1)(0,1)\pr\pr}+\0{2}{r}v_2^{(1)(0,1)\pr}\left(1+\0{r h_{G+1}'}{h_{G+1}}\right)+\0{2(2G+1)}{r^2}v_2^{(1)(0,1)} \\
& \qquad -\0{\pi}{r}\0{2\sqrt{G(G+1)}}{2G+1}+\0{2\sqrt{G(G+1)}}{2G+1}\pi '=0,
\end{aligned}\right. \\
&\left\{\begin{aligned}
& v_1^{(2)(0,1)\pr\pr}+\0{2}{r}v_1^{(2)(0,1)\pr}\left(1+\0{r h_{G-1}^\pr}{h_{G-1}}\right)-\0{2(2G+1)}{r^2}v_1^{(2)(0,1)} \\
& \qquad -\0{\pi}{r}\0{2\sqrt{G(G+1)}}{2G+1}+\0{2\sqrt{G(G+1)}}{2G+1}\pi '=0, \\
& v_2^{(2)(0,1)\pr\pr}+\0{2}{r}v_2^{(2)(0,1)\pr}\left(1+\0{r h_{G-1}^\pr}{h_{G-1}}\right)+\0{G}{2G+1}\0{2}{r}\pi+\0{\pi '}{2G+1}=0,
\end{aligned}\right.
\end{align}

\item $g_p^2$ order equations:
\hspace{-5cm}
\begin{align}
&\left\{\begin{aligned}
& \left(-\0{\pi '}{2G+1}+\0{2}{r}\0{G+1}{2G+1}\pi\right)v_1^{(1)(0,1)}+\0{2\sqrt{G(G+1)}}{2G+1}\left(\pi '-\0{\pi}{r}\right)v_2^{(1)(0,1)} \\
& \qquad -\pi^2+v_1^{(1)(0,2)\pr\pr}+\0{2}{r}v_1^{(1)(0,2)\pr}\left(1+\0{r h_{G+1}'}{h_{G+1}}\right)=0, \\
& \0{2\sqrt{G(G+1)}}{2G+1}\left(\pi '-\0{\pi}{r}\right)v_1^{(1)(0,1)}+\left(\0{\pi '}{2G+1}+\0{2}{r}\0{G}{2G+1}\pi\right)v_2^{(1)(0,1)} \\
& \qquad +v_2^{(1)(0,2)\pr\pr}+\0{2}{r}v_2^{(1)(0,2)\pr}\left(1+\0{r h_{G+1}'}{h_{G+1}}\right)+\0{2(2G+1)}{r^2}v_2^{(1)(0,2)}=0,
\end{aligned}\right. \\
&\left\{\begin{aligned}
& \left(-\0{\pi '}{2G+1}+\0{2}{r}\0{G+1}{2G+1}\pi\right)v_1^{(2)(0,1)}+\0{2\sqrt{G(G+1)}}{2G+1}\left(\pi '-\0{\pi}{r}\right)v_2^{(2)(0,1)} \\
& \qquad +v_1^{(2)(0,2)\pr\pr}+\0{2}{r}v_1^{(2)(0,2)\pr}\left(1+\0{r h_{G-1}'}{h_{G-1}}\right)-\0{2(2G+1)}{r^2}v_1^{(2)(0,2)}=0, \\
& \0{2\sqrt{G(G+1)}}{2G+1}\left(\pi '-\0{\pi}{r}\right)v_1^{(2)(0,1)}+\left(\0{\pi '}{2G+1}+\0{2}{r}\0{G}{2G+1}\pi\right)v_2^{(2)(0,1)}-\pi^2 \\
& \qquad +v_2^{(2)(0,2)\pr\pr}+\0{2}{r}v_2^{(2)(0,2)\pr}\left(1+\0{r h_{G-1}'}{h_{G-1}}\right)=0,
\end{aligned}\right.
\end{align}

\item $g_sg_p$ order equations:
\hspace{-5cm}
\begin{align}
&\left\{\begin{aligned}
& \left(-\0{\pi '}{2G+1}+\0{2}{r}\0{G+1}{2G+1}\pi\right)v_1^{(1)(0,1)}+\0{2\sqrt{G(G+1)}}{2G+1}\left(\pi '-\0{\pi}{r}\right)v_2^{(1)(1,0)} \\
& \qquad +\0{\s_{s} '}{\e_v}\0{\pi}{2G+1}-\0{\s_{s} '}{\e_v}\left(v_1^{(1)(0,1)\pr}+v_1^{(1)(0,1)}\0{h_{G+1}'}{h_{G+1}}\right)-\0{\s_{s} '}{\e_v}\0{G+2}{r}v_1^{(1)(0,1)} \\
& \qquad +v_1^{(1)(1,1)\pr\pr}+\0{2}{r}v_1^{(1)(1,1)\pr}\left(1+\0{r h_{G+1}'}{h_{G+1}}\right)=0, \\
& \0{2\sqrt{G(G+1)}}{2G+1}\left(\pi '-\0{\pi}{r}\right)v_1^{(1)(1,0)}+\left(\0{\pi '}{2G+1}+\0{2}{r}\0{G}{2G+1}\pi\right)v_2^{(1)(1,0)} \\
& \qquad -\0{\s_{s} '\pi}{\e_v}\0{2\sqrt{G(G+1)}}{2G+1}-\0{\s_{s} '}{\e_v}\left(v_2^{(1)(0,1)\pr}+v_2^{(1)(0,1)}\0{h_{G+1}'}{h_{G+1}}\right) \\
& \qquad +\0{\s_{s} '}{\e_v}\0{G-1}{r}v_2^{(1)(0,1)}+v_2^{(1)(1,1)\pr\pr}+\0{2}{r}v_2^{(1)(1,1)\pr}\left(1+\0{r h_{G+1}'}{h_{G+1}}\right) \\
& \qquad +\0{2(2G+1)}{r^2}v_2^{(1)(1,1)}=0,
\end{aligned}\right. \\
&\left\{\begin{aligned}
& \left(-\0{\pi '}{2G+1}+\0{2}{r}\0{G+1}{2G+1}\pi\right)v_1^{(2)(1,0)}+\0{2\sqrt{G(G+1)}}{2G+1}\left(\pi '-\0{\pi}{r}\right)v_2^{(2)(1,0)} \\
& \qquad -\0{\s_{s} '\pi}{\e_v}\0{2\sqrt{G(G+1)}}{2G+1}-\0{\s_{s} '}{\e_v}\left(v_1^{(2)(0,1)\pr}+v_1^{(2)(0,1)}\0{h_{G-1}'}{h_{G-1}}\right) \\
& \qquad -\0{\s_{s} '}{\e_v}\0{G+2}{r}v_1^{(2)(0,1)}+v_1^{(2)(1,1)\pr\pr}+\0{2}{r}v_1^{(2)(1,1)\pr}\left(1+\0{r h_{G-1}'}{h_{G-1}}\right) \\
& \qquad -\0{2(2G+1)}{r^2}v_1^{(2)(1,1)}=0, \\
& \0{2\sqrt{G(G+1)}}{2G+1}\left(\pi '-\0{\pi}{r}\right)v_1^{(2)(1,0)}+\left(\0{\pi '}{2G+1}+\0{2}{r}\0{G}{2G+1}\pi\right)v_2^{(2)(1,0)} \\
& \qquad -\0{\s_{s} '}{\e_v}\0{\pi}{2G+1}-\0{\s_{s} '}{\e_v}\left(v_2^{(2)(0,1)\pr}+v_2^{(2)(0,1)}\0{h_{G-1}'}{h_{G-1}}\right)+\0{\s_{s} '}{\e_v}\0{G-1}{r}v_2^{(2)(0,1)} \\
& \qquad +v_2^{(2)(1,1)\pr\pr}+\0{2}{r}v_2^{(2)(1,1)\pr}\left(1+\0{r h_{G-1}'}{h_{G-1}}\right)=0,
\end{aligned}\right.
\end{align}

\end{enumerate}
where $\varepsilon_{v} =\varepsilon+g\sigma_v$, $\sigma_{s}=\tilde{\sigma}(r)$ and $\pi=\tilde{\pi}(r)$. The above equations could be numerically solved order by order. Once the elements $v_i^{(j)(n_s,\,n_p)}$ of different orders are obtained, the different orders of Born subtractions of the phase shifts could be evaluated in the same way as shown in Appendix B. We have verified that the phase shifts evaluated by the second order differential equations to the fourth order give the identical results as those evaluated by the first order differential equations.

\section{Comparisons between results from Born subtraction and fake boson methods}
In this appendix, we make a comparison between the results of the total energies from Born subtraction and fake boson methods. According to Ref.\cite{farhi2002searching}, the total energy of the assumed background field
configuration with fermion number $1$ is
\begin{equation}
	\frac{1}{m}E=
	{\cal E}_{\rm cl}(w,b_1,b_2,c)
	+\left(1-Q_{\rm vac}\right)
	\epsilon_1(w,b_1,b_2,c)+{\cal E}_{\rm vac}(w,b_1,b_2,c),
	\label{etot}
\end{equation}
with a four parameters ansatz of the background fields in the following form
\begin{eqnarray}
s + i \vec{\tau} \cdot \vec{p} = \rho(\xi) \exp\left(i \vec{\tau} \cdot \hat{\vec{r}} \Theta(\xi)\right), \\
\rho(\xi) = 1 + b_1 \left[ 1 + b_2^2 \frac{\xi}{w} \right] \exp\left(-b_2^2 \frac{\xi}{w}\right), \\
\Theta(\xi) = -\pi \frac{e^{c^2} - 1}{e^{c^2} - 3 + 2e^{c^2 \xi / w}},
\end{eqnarray}
where $m$ is the fermion mass $m=g\s_v$. $s$ and $\vec p$ are the dimensionless fields which can be identified as $s=\s/\s_v$ and $\vec p=\vec\pi/\s_v$ in our case. All the energies and lengths are scaled in terms of the fermion mass $m$, which means $\xi=mr$, ${\cal E}_{\rm cl}=E_{\rm cl}/m$, ${\cal E}_{\rm vac}=E_{\rm vac}/m$ and $\epsilon_1=E_1/m$. It should be noted that $E_1$ is the energy eigenvalue of the most strongly bound state. $E_{\rm cl}$ is the classical energy of the background fields. The background field could be determined by the four variational parameters $w$, $b_1$, $b_2$ and $c$.
The polarized vacuum charge $Q_{\rm vac}$ is given by
\begin{eqnarray}
	Q_{\rm vac}&=&\sum_{G,\Pi}(2G+1)\left[
	\frac{1}{\pi}\delta_{G,+,\Pi}(0)-n^{(+)}_{G,\Pi}\right]
	\nonumber \\
	&=&-\sum_{G,\Pi}(2G+1)\left[
	\frac{1}{\pi}\delta_{G,-,\Pi}(0)-n^{(-)}_{G,\Pi}\right],
	\label{qvac}
\end{eqnarray}
where $n^{(+)}_{G,\Pi}$ and $n^{(-)}_{G,\Pi}$ are the numbers of positive and negative energy bound states, respectively.

\begin{figure}[htbp]
	\centering
	\hspace{-1.5cm}
	\begin{minipage}{0.8\textwidth}
		\centering
		\includegraphics[width=\textwidth]{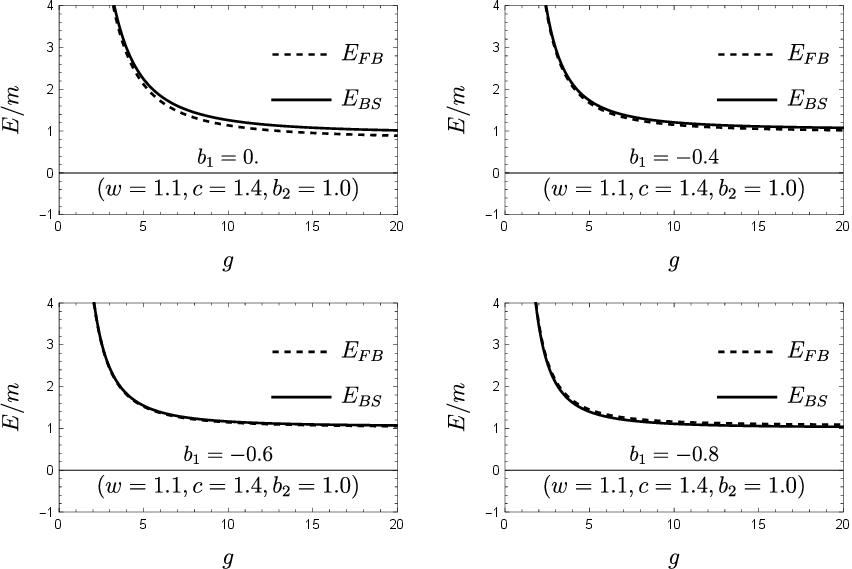}
	\end{minipage}\\
	\caption{The comparison of the scaled total energies as functions of the coupling constant $g$ calculated by the Born subtraction and the fake boson methods at different sets of parameters.}\label{b1-4}
\end{figure}

In order to compare the results from the Born subtraction (BS) and the fake boson (FB) methods,  we have selected several sample sets of the parameters from Ref.\cite{farhi2002searching} to evaluate the total energy. The results are show in Fig.\ref{b1-4}. The solid and dashed lines show the scaled total energies as functions of the coupling constant $g$, calculated by the BS and FB methods, respectively. It could be seen that the two results agree reasonably well with each other at different sets of parameters. This can be viewed as a mutual validation of the two computational methods.

\clearpage
\bibliographystyle{apsrev}
\bibliography{ref11}

@article{PhysRevD.80.074025,
  title = {Inhomogeneous phases in the Nambu--Jona-Lasinio and quark-meson model},
  author = {Nickel, Dominik},
  journal = {Phys. Rev. D},
  volume = {80},
  issue = {7},
  pages = {074025},
  numpages = {13},
  year = {2009},
  month = {Oct},
  publisher = {American Physical Society}
}

@article{FUKUSHIMA201399,
title = {The phase diagram of nuclear and quark matter at high baryon density},
journal = {Progress in Particle and Nuclear Physics},
volume = {72},
pages = {99-154},
year = {2013},
author = {Kenji Fukushima and Chihiro Sasaki}
}

@article{BUBALLA201539,
title = {Inhomogeneous chiral condensates},
journal = {Progress in Particle and Nuclear Physics},
volume = {81},
pages = {39-96},
year = {2015},
author = {Michael Buballa and Stefano Carignano}
}

@article{Brauner2017csl,
  title={Chiral soliton lattice and charged pion condensation in strong magnetic fields},
  author={Brauner, Tom{\'a}{\v{s}} and Yamamoto, Naoki},
  journal={Journal of High Energy Physics},
  volume={2017},
  number={4},
  pages={1--19},
  year={2017},
  publisher={Springer}
}

@article{chen2021,
  title={Inhomogeneous confining-deconfining phases in rotating plasmas},
  author={Chernodub, Maxim N},
  journal={Physical Review D},
  volume={103},
  number={5},
  pages={054027},
  year={2021},
  publisher={APS}
}

@article{chen2022,
  title={Inhomogeneous chiral condensation under rotation in the holographic QCD},
  author={Chen, Yidian and Li, Danning and Huang, Mei},
  journal={Physical Review D},
  volume={106},
  number={10},
  pages={106002},
  year={2022},
  publisher={APS}
}

@article{Ishioka2010cdw,
  title={Chiral charge-density waves},
  author={Ishioka, J and Liu, YH and Shimatake, K and Kurosawa, T and Ichimura, K and Toda, Y and Oda, M and Tanda, S},
  journal={Physical review letters},
  volume={105},
  number={17},
  pages={176401},
  year={2010},
  publisher={APS}
}

@article{VOSKRESENSKY2023104030,
title = {Structure formation during phase transitions in strongly interacting matter},
journal = {Progress in Particle and Nuclear Physics},
volume = {130},
pages = {104030},
year = {2023},
author = {D.N. Voskresensky}
}

@article{PhysRevD.103.034030,
  title = {Competition of inhomogeneous chiral phases and two-flavor color superconductivity in the NJL model},
  author = {Lakaschus, Phillip and Buballa, Michael and Rischke, Dirk H.},
  journal = {Phys. Rev. D},
  volume = {103},
  issue = {3},
  pages = {034030},
  numpages = {9},
  year = {2021},
  month = {Feb}
}

@article{PhysRevLett.127.232002,
  title = {Spontaneous Symmetry Breaking via Inhomogeneities and the Differential Surface Tension},
  author = {Endr\ifmmode \mbox{\H{o}}\else \H{o}\fi{}di, G. and Kov\'acs, T. G. and Mark\'o, G.},
  journal = {Phys. Rev. Lett.},
  volume = {127},
  issue = {23},
  pages = {232002},
  numpages = {6},
  year = {2021},
  month = {Dec},
  publisher = {American Physical Society}
}

@article{Cardall_2001,
year = {2001},
month = {jun},
publisher = {},
volume = {554},
number = {1},
pages = {322},
author = {Cardall, Christian Y. and Prakash, Madappa and Lattimer, James M.},
title = {Effects of Strong Magnetic Fields on Neutron Star
Structure},
journal = {The Astrophysical Journal}
}

@article{PhysRevLett.128.061101,
  title = {Magnetohydrodynamic Simulations of Self-Consistent Rotating Neutron Stars with Mixed Poloidal and Toroidal Magnetic Fields},
  author = {Tsokaros, Antonios and Ruiz, Milton and Shapiro, Stuart L. and Ury\ifmmode \bar{u}\else \={u}\fi{}, K\ifmmode \bar{o}\else \={o}\fi{}ji},
  journal = {Phys. Rev. Lett.},
  volume = {128},
  issue = {6},
  pages = {061101},
  numpages = {7},
  year = {2022},
  month = {Feb},
  publisher = {American Physical Society}
}

@article{BZDAK20201,
title = {Mapping the phases of quantum chromodynamics with beam energy scan},
journal = {Physics Reports},
volume = {853},
pages = {1-87},
year = {2020},
author = {Adam Bzdak and ShinIchi Esumi and Volker Koch and Jinfeng Liao and Mikhail Stephanov and Nu Xu}
}

@article{PhysRevD.109.036023,
  title = {Thermal phonon fluctuations and stability of the magnetic dual chiral density wave phase in dense QCD},
  author = {Ferrer, E. J. and Gyory, W. and de la Incera, V.},
  journal = {Phys. Rev. D},
  volume = {109},
  issue = {3},
  pages = {036023},
  numpages = {14},
  year = {2024},
  month = {Feb},
  publisher = {American Physical Society}
}

@article{Evans2012baryonic,
  title={The baryonic phase in holographic descriptions of the QCD phase diagram},
  author={Evans, Nick and Kim, Keun-Young and Magou, Maria and Seo, Yunseok and Sin, Sang-Jin},
  journal={Journal of High Energy Physics},
  volume={2012},
  number={9},
  pages={1--23},
  year={2012},
  publisher={Springer}
}

@article{PhysRevD.37.1670,
  title = {Calculating boson and fermion loops in 3+1 dimensions and the derivative expansion},
  author = {Li, Ming and Perry, Robert J.},
  journal = {Phys. Rev. D},
  volume = {37},
  issue = {6},
  pages = {1670--1676},
  numpages = {0},
  year = {1988},
  month = {Mar},
  publisher = {American Physical Society}
}

@article{PhysRevD.55.3742,
  title = {Quantum solitons at strong coupling},
  author = {Stewart, I. W. and Blunden, P. G.},
  journal = {Phys. Rev. D},
  volume = {55},
  issue = {6},
  pages = {3742--3747},
  numpages = {0},
  year = {1997},
  month = {Mar},
  publisher = {American Physical Society}
}

@article{KAHANA1984462,
title = {Baryon density of quarks coupled to a chiral field},
journal = {Nuclear Physics A},
volume = {429},
number = {3},
pages = {462-476},
year = {1984},
author = {S. Kahana and G. Ripka}
}

@article{Rebhan2005BPSSO,
  title={BPS saturation of the N=4 monopole by infinite composite-operator renormalization},
  author={Anton Rebhan and R. Schofbeck and Peter van Nieuwenhuizen and Robert Wimmer},
  journal={Physics Letters B},
  year={2005},
  volume={632},
  pages={145-150}
}

@article{PhysRevD.59.045016,
  title = {Anomaly and quantum corrections to solitons in two-dimensional theories with minimal supersymmetry},
  author = {Shifman, M. and Vainshtein, A. and Voloshin, M.},
  journal = {Phys. Rev. D},
  volume = {59},
  issue = {4},
  pages = {045016},
  numpages = {25},
  year = {1999},
  month = {Jan},
  publisher = {American Physical Society}
}

@article{DUNNE1999238,
title = {Derivative expansion and soliton masses},
journal = {Physics Letters B},
volume = {467},
number = {3},
pages = {238-246},
year = {1999},
author = {Gerald V. Dunne}
}

@article{REBHAN2004234,
title = {A new anomalous contribution to the central charge of the N=2 monopole},
journal = {Physics Letters B},
volume = {594},
number = {1},
pages = {234-240},
year = {2004},
author = {A. Rebhan and P. {van Nieuwenhuizen} and R. Wimmer}
}

@article{GRAHAM1999432,
title = {Energy, central charge, and the BPS bound for 1 + 1-dimensional supersymmetric solitons},
journal = {Nuclear Physics B},
volume = {544},
number = {1},
pages = {432-447},
year = {1999},
author = {N. Graham and R.L. Jaffe}
}

@book{Graham:2009zz,
    author = "Graham, Noah and Quandt, Markus and Weigel, Herbert",
    title = "{Spectral Methods in Quantum Field Theory}",
    doi = "10.1007/978-3-642-00139-0",
    volume = "777",
    month = "4",
    year = "2009"
}

@article{farhi1998finite,
  title={Finite quantum fluctuations about static field configurations},
  author={Farhi, Edward and Graham, Noah and Haagensen, P and Jaffe, Robert L},
  journal={Physics Letters B},
  volume={427},
  number={3-4},
  pages={334--342},
  year={1998},
  publisher={Elsevier}
}

@article{farhi2002searching,
  title={Searching for quantum solitons in a (3+ 1)-dimensional chiral Yukawa model},
  author={Farhi, E and Graham, N and Jaffe, RL and Weigel, H},
  journal={Nuclear Physics B},
  volume={630},
  number={1-2},
  pages={241--268},
  year={2002},
  publisher={Elsevier}
}

@article{cosmicstring,
title = {Quantum stabilization of a hedgehog type of cosmic string},
author = {M. Quandt and N. Graham and H. Weigel},
journal = {Nuclear Physics B},
volume = {923},
pages = {350-377},
year = {2017},
publisher={Elsevier}
}

@article{ma_proof_1985,
  title = {Proof of the {Levinson} theorem by the {Sturm}–{Liouville} theorem},
  journal = {Journal of Mathematical Physics},
  author = {Ma, Zhong‐Qi},
  volume = {26},
  number = {8},
  pages = {1995--1999},
  year = {1985},
  publisher={AIP}	
}

@article{schwinger1954theory,
  title={The theory of quantized fields. VI},
  author={Schwinger, Julian},
  journal={Physical Review},
  volume={94},
  number={5},
  pages={1362},
  year={1954},
  publisher={APS}
}

@article{tHooft1976instanton,
  title={Computation of the quantum effects due to a four-dimensional pseudoparticle},
  author={t Hooft, Gerard},
  journal={Physical Review D},
  volume={14},
  number={12},
  pages={3432--3450},
  year={1976},
  publisher={Elsevier}
}

@article{SHU2017soliton,
  author = {Song Shu},
  title = {A practical method in calculating one loop quantum fluctuations to the energy of the non-topological soliton},
  journal = {SCIENCE CHINA Physics, Mechanics {\&} Astronomy},
  year = {2017},
  volume = {60},
  number = {4},
  pages = {041021-}
}

@article{SHU2021122256,
title = {Vacuum and thermal fluctuation energies of a soliton at finite temperatures},
journal = {Nuclear Physics A},
volume = {1014},
pages = {122256},
year = {2021},
author = {Song Shu and Xiaogang Li and Jia-Rong Li}
}

@article{li2022fl,
  title={Quantum fluctuation in an inhomogeneous background and its influence on the phase transition in a finite volume system},
  author={Li, Xiaogang and Shu, Song and Li, Jia-Rong},
  journal={Physical Review C},
  volume={105},
  number={4},
  pages={045203},
  year={2022},
  publisher={APS}
}

@article{BIRSE1984284,
title = {A chiral soliton model of nucleon and delta},
journal = {Physics Letters B},
volume = {136},
number = {4},
pages = {284-288},
year = {1984},
author = {Michael C. Birse and Manoj K. Banerjee}
}

@article{diakonov1988chiral,
  title={A chiral theory of nucleons},
  author={Diakonov, DI and Petrov, V Yu and Pobylitsa, PV},
  journal={Nuclear Physics B},
  volume={306},
  number={4},
  pages={809--848},
  year={1988},
  publisher={Elsevier}
}

@article{christov1996baryons,
  title={Baryons as non-topological chiral solitons},
  author={Christov, Chr V and Blotz, A and Kim, H-C and Pobylitsa, P and Watabe, T and Meissner, Th and Arriola, E Ruiz and Goeke, K},
  journal={Progress in Particle and Nuclear Physics},
  volume={37},
  pages={91--191},
  year={1996},
  publisher={Elsevier}
}

@article{PhysRevC.70.015204,
  title = {Role of fluctuations in the linear $\ensuremath{\sigma}$ model with quarks},
  author = {M\'ocsy, \'A. and Mishustin, I. N. and Ellis, P. J.},
  journal = {Phys. Rev. C},
  volume = {70},
  issue = {1},
  pages = {015204},
  numpages = {11},
  year = {2004},
  month = {Jul},
  publisher = {American Physical Society}
}

@article{WENG2025139587,
title = {The nucleon properties in finite temperature and density with Gaussian fluctuations},
journal = {Physics Letters B},
volume = {867},
pages = {139587},
year = {2025},
author = {Peixin Weng and Bingtao Li and Yiming Lyu and Song Shu and Hui Zhang}
}

\end{document}